\documentclass[confeence, final, twocolumn,10pt]{IEEEtran}
\usepackage{cite}
\usepackage{graphicx}
\usepackage{subfigure}
\usepackage{mathrsfs}
\usepackage{amsmath}
\usepackage{amsfonts}
\usepackage{amssymb}
\usepackage{textcomp}

\allowdisplaybreaks

\newcommand{\qed}{\hspace*{\fill} $\Box$ \\}
\newcommand{\bs}{\boldsymbol}

\def\ba{\begin{array}}
\def\ea{\end{array}}
\def\be{\begin{equation}}
\def\ee{\end{equation}}
\def\bea{\begin{eqnarray}}
\def\eea{\end{eqnarray}}
\def\beas{\begin{eqnarray*}}
\def\eeas{\end{eqnarray*}}

\newtheorem{theorem}{Theorem}

\newtheorem{lemma}{Lemma}
\newtheorem{proposition}{Proposition}

\title{Distributed Algorithms for Spectrum Allocation, Power Control, Routing, and Congestion Control in Wireless Networks}
\author{\authorblockN{Yufang Xi and Edmund M. Yeh}\\
\authorblockA{Department of Electrical Engineering\\
Yale University\\
New Haven, CT 06520, USA \\ Email: \{yufang.xi, edmund.yeh\}@yale.edu}}

\begin{document}

\maketitle

\setcounter{footnote}{1} \footnotetext{This research is supported in part by NSF grant
CNS-0626882 and AFOSR grant FA9550-06-1-0135.}

\begin{abstract}

We develop distributed algorithms to allocate resources in multi-hop wireless networks
with the aim of minimizing total cost. In order to observe the fundamental duplexing
constraint that co-located transmitters and receivers cannot operate simultaneously on
the same frequency band, we first devise a spectrum allocation scheme that divides the
whole spectrum into multiple sub-bands and activates conflict-free links on each
sub-band.  We show that the minimum number of required sub-bands grows asymptotically at
a logarithmic rate with the chromatic number of network connectivity graph. A simple
distributed and asynchronous algorithm is developed to feasibly activate links on the
available sub-bands. Given a feasible spectrum allocation, we then design node-based
distributed algorithms for optimally controlling the transmission powers on active links
for each sub-band, jointly with traffic routes and user input rates in response to
channel states and traffic demands. We show that under specified conditions, the
algorithms asymptotically converge to the optimal operating point.

\end{abstract}


\section{\label{sec:Introduction}Introduction}

While offering the potential for ubiquitous and untethered communications, wireless
networks typically demand more sophisticated resource management than wireline networks.
Optimal resource allocation in large-scale wireless networks involves joint optimization
across multiple layers as well as distributed implementation across network nodes. In
this paper, we develop distributed algorithms which jointly allocate frequency spectrum,
transmission powers, traffic input rates, and traffic routes on a node-by-node basis in
order to minimize total cost in an interference-limited multi-hop wireless network.

While joint optimization involving power control, congestion control, and routing has
been studied previously~\cite{paper:JXB03,paper:Chi04,paper:XY06ISIT1}, a common
shortcoming of the previous work is the failure to account for the constraint that a node
cannot transmit and receive simultaneously on the same frequency band.  For while some
types of multi-user interference can be ameliorated by using advanced coding techniques,
interference between a transmitter and a co-located receiver is very difficult to
suppress due to the transmitted power being many orders of magnitude higher than the
received power at the same node.\footnote{Theoretically a node is able to subtract the
transmission signals generated by itself from the received signals so that the
self-interference can be perfectly cancelled out. In real decoders, however, the received
signals are practically irrecoverable in the face of the overwhelming transmitted
signals.} The resulting constraint that a transmitter cannot be simultaneously active
with a co-located receiver~\cite{book:Gol05} on the same frequency band is referred to as
the {\em duplexing constraint}. In practical wireless networks, the duplexing constraint
appears to be quite fundamental, and thus must be observed by resource management
schemes.

In this paper, we develop distributed resource allocation algorithms for wireless
networks in accordance with the duplexing constraint.  To accomplish this, we first
devise a new spectrum allocation scheme which divides the spectrum into a sufficient
number of frequency bands and activates co-located transmitters and receivers on
different bands.  We show that the minimum number of sub-bands needed for resolving
duplexing conflicts is asymptotically {\em logarithmic} in the chromatic
number\footnote{The chromatic number of a graph is the minimum number of colors with
which the vertices of the graph can be colored such that adjacent vertices have different
colors.} of the network connectivity graph. We provide a simple algorithm that feasibly
assigns frequency bands to links in a distributed and asynchronous manner with low
control overhead.

Given a conflict-free spectrum allocation, we then design a set of node-based distributed
gradient projection algorithms that iteratively adjust transmission powers, traffic input
rates, and traffic routes according to channel conditions and traffic demands, in order
to minimize total network cost.  The power control and routing algorithms we develop are
frequency selective, in that for each link, the power control algorithm adjusts the
transmission power on each of the link's active sub-bands. The routing algorithm involves
both {\em inter-node routing}, which specifies the allocation of incoming traffic at each
node to its outgoing links, and {\em intra-node routing}, which specifies the allocation
of the total traffic on a given link across its active sub-bands.  Finally, we show that
congestion control can be naturally incorporated by considering an equivalent routing
problem on a virtual overflow link.  We show that under specified conditions, the
iterative algorithms converge to the optimal operating point from any initial condition.

The rest of the paper is organized as follows.  In Section~\ref{sec:Model}, we introduce
the network model, discuss the duplexing constraint, and formulate the spectrum
allocation problem. In Section~\ref{sec:MinColor}, we find the minimum number of
sub-bands required by a feasible spectrum allocation. A distributed and asynchronous
sub-band allocation algorithm is developed in Section~\ref{sec:Subband}. In
Section~\ref{sec:Optimization}, we formulate the cross-layer optimization problem for
networks operating on multiple sub-bands, and derive the conditions satisfied by the
optimal configuration. In Section~\ref{sec:Algorithms}, we present node-based gradient
projection algorithms to jointly optimize power control, routing, and congestion control
based on the outcome of the spectrum allocation, as well as the channel conditions and
traffic demands of the network.

\section{\label{sec:Model}Network Model and Spectrum Allocation}

Let the wireless network be modelled by a directed and connected graph $\mathcal G = (
{\cal N}, {\cal L} )$, where $\mathcal G$ is referred to as the {\em connectivity graph}
of the network. A node $i\in {\cal N}$ represents a wireless transceiver, and $(i,j) \in
{\cal L}$ represents a unidirectional wireless link from node $i$ to $j$. We assume
$\mathcal G$ is link-symmetric, i.e., if $(i,j)\in {\cal L}$, then $(j,i)\in {\cal L}$,
and vice versa. Two nodes $i$ and $j$ are {\em neighbors} if $(i,j) \in {\cal L}$ (or
equivalently $(j,i)\in {\cal L}$). Let ${\cal O}_i \triangleq \{j: (i,j)\in {\cal L}\}$
be the set of $i$'s neighbors.

\subsection{Duplexing Constrained Interference Model}

We focus on a network model that prohibits any node from simultaneously transmitting and
receiving on the same frequency band. That is, we impose the duplexing constraint on
every band.  Such a constraint is less stringent but more fundamental than the
extensively studied primary interference
constraint~\cite{paper:CLC06,paper:BES06,paper:HS88}, where any node can transmit or
receive (at any given time and on any given band) on at most one active link, and the
secondary interference constraint, which further prohibits any node from transmitting
when there is a neighbor receiving from another node~\cite{paper:MSZ06}. A general
approach to studying a broad class of interference constraints was presented
in~\cite{paper:Ram97}.
Indeed, duplexing constraints cannot be bypassed (at least currently) by using
sophisticated coding methods, and must be observed by practical network management
techniques.

Traditional network management techniques which aim to resolve various types of
interference among wireless links have concentrated on scheduling in
time~\cite{tass92,paper:MAW96,pat:Tse,
paper:SA02,paper:LCS01,paper:LK03,paper:Bor03,paper:ElE02}, where at any given time, only
mutually-non-interfering links are activated. Scheduling, however, usually requires
centralized controllers and involves high communication and computational
complexity~\cite{paper:HS88,paper:Ari84}. Simplified distributed scheduling policies have
been proposed~\cite{paper:NMR02,paper:MSZ06,paper:CKS05,paper:CLC06,paper:BES06} for
various purposes. In general, however, the reduced implementation complexity comes at the
expense of performance~\cite{paper:LS05}.

The difficulty in finding interference-free schedules in time leads us to seek an
alternative solution.  A natural approach is to consider network management in the domain
of {\em frequency} instead of or in addition to
time~\cite{paper:Hal80,paper:CW84,paper:ABL05,paper:KN05}. Because communication on
different frequency bands are practically non-interfering, one can think of
simultaneously applying different link activation sets on non-overlapping frequency bands
within the assigned spectrum.   In this scheme, nodes transmit on certain bands while
receiving on other bands to avoid duplexing interference. The spectrum allocation
technique has an important advantage over scheduling in time: once a feasible spectrum
allocation is established, the network can function relatively statically in that mode,
not having to switch to another mode unless network itself changes substantially. While
spectrum allocation problems have been proposed and studied in the interference graph
induced by particular interference constraints, the solution using existing
vertex-coloring methods~\cite{paper:Hal80,paper:CW84} is cumbersome and its complexity
scales poorly with the size of the network.  Moreover, the optimization of frequency
assignment techniques has not been thoroughly investigated.  In particular, the number of
available frequency bands is often arbitrarily set and frequency bands are assigned to
links in a heuristic manner \cite{paper:ABL05,paper:KN05}.

In this work, we adopt the spectrum allocation approach to resolve the fundamental
duplexing conflicts for general wireless networks. In particular, we investigate two
central questions: (1) what is the minimum number of frequency bands with which all
co-located transmitters and receivers can be simultaneously activated subject to
duplexing constraints, and (2) how can one efficiently find a feasible frequency
assignment when there are enough frequency bands? We provide an exact analytical answer
to question (1) and develop a distributed asynchronous algorithm which solves problem
(2). Our analysis is based only the network connectivity graph. This approach requires
much less storage and computation overhead than alternative methods that utilize the
interference graph.

\subsection{Spectrum Division and Sub-band Allocation}

The duplexing constraint permits only a subset of the links to be activated
simultaneously on each frequency band. To activate all links at the same time, it is
necessary to divide the spectrum into several sub-bands and activate different subsets of
conflict-free links on different sub-bands.

Suppose the network occupies a contiguous spectrum which can be partitioned into a
number, say $Q$, of sub-bands, each of which covers a contiguous segment of the whole
spectrum. Let the collection of the sub-bands be denoted by ${\cal Q}$. With a specific
spectrum division in place, each link can be active on one or more of the sub-bands. If
link $(i,j)$ uses sub-band $q$, we say $(i,j)$ is an {\em active link} on $q$, and $q$ is
an {\em active sub-band} of $(i,j)$. Denote the subset of links that are active on a
sub-band $q$ by ${\cal L}_q$, and the set of active sub-bands of $(i,j)$ by ${\cal
Q}_{ij}$.  A {\em spectrum allocation} is given by the collection $\{{\cal
L}_q\}_{q\in{\cal Q}}$ (or equivalently $\{{\cal Q}_{ij}\}_{(i,j) \in {\cal L}}$). Note
that finding a spectrum allocation involves two steps: {\em spectrum division}, which
decides how many sub-bands the whole spectrum is divided into, and {\em sub-band
allocation}, which determines which links are active on which sub-bands. We will address
these two issues in Sections~\ref{sec:MinColor} and \ref{sec:Subband}, respectively.

A spectrum allocation is {\em feasible} if (i) for all $(i,j)\in{\cal L}$, ${\cal Q}_{ij}
\ne \emptyset$, and (ii) for all $q\in{\cal Q}$, ${\cal L}_q$ satisfies the duplexing
constraint. Thus, any node's outgoing and incoming links cannot be both active on the
same sub-band. However, it is feasible for a node to have multiple active outgoing links
or multiple active incoming links on a single sub-band.

\subsection{Interference Graph and Number of Sub-bands}\label{subsec:IntApp}

%

\begin{figure*}[!t]
\begin{center}
  \includegraphics[width=15cm]{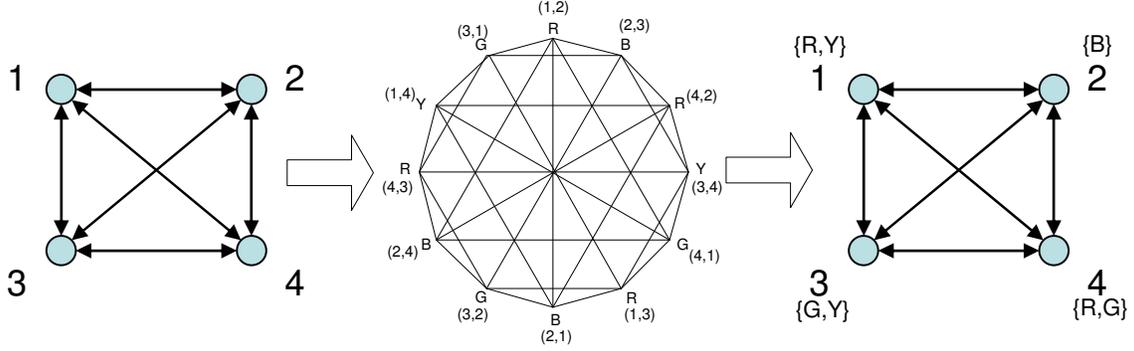}\\
  \caption{The interference graph approach on the complete graph with four nodes}\label{fig:IntApp}
\end{center}
\end{figure*}

In previous studies on frequency assignment techniques, the minimum number of frequency
bands is found from the interference graph induced by the specific interference
constraints~\cite{paper:Hal80,paper:CW84}. 
%
For the duplexing constraints, the interference graph $\tilde{\cal G}$ is constructed as
follows. Let the vertices of the interference graph $\tilde{\cal G} = ({\cal V}, {\cal
E})$ be the {\em links} of the network connectivity graph ${\cal G} = ({\cal N}, {\cal
L})$, i.e., ${\cal V} = {\cal L}$. In $\tilde{\cal G}$, an edge exists between two
vertices (links in ${\cal G}$) if one link's transmitter is the other link's receiver.
The interference graph, unlike the connectivity graph, is undirected.\footnote{In order
to avoid confusion, we refer to the connectivity graph as having nodes ${\cal N}$ and
links ${\cal L}$, and the interference graph as having vertices ${\cal V}$ and edges
${\cal E}$. }
%
%
It is easy to see that a feasible spectrum allocation on ${\cal G}$ exists if and only if
the number of available sub-bands is greater than or equal to the chromatic number
$\chi(\tilde{\cal G})$ of the interference graph $\tilde{\cal G}$. We illustrate the
interference graph approach in Figure~\ref{fig:IntApp} using a complete four-node
network. On the left is the connectivity graph ${\cal G}$, whose induced interference
graph $\tilde{\cal G}$ is shown in the middle. A minimal vertex-coloring on $\tilde{\cal
G}$ using four colors $\{R,G,B,Y\}$ is depicted. The right graph represents the
link-coloring on ${\cal G}$ inferred by the minimal vertex-coloring on $\tilde{\cal G}$.
The color(s) associated with each node are those assigned to the outgoing links of that
node. The reason for using such a representation will be explained shortly.

Although the interference graph approach can provide an answer to our problem, it has
major shortcomings.  Note that the size of the interference graph $|{\cal V}| = |{\cal
L}|$ is on the order of $|{\cal N}|^2$.  To compute the chromatic number of $\tilde{\cal
G}$, $\tilde{\cal G}$ has to be constructed and stored at a central controller which then
computes $\chi(\tilde{\cal G})$ by finding a minimal vertex-coloring, which is itself an
NP-complete problem~\cite{book:GJ79}.  For these reasons, the interference graph approach
is not tractable for medium-and-large-scale networks.  This motivates us to seek an
alternative method that can find the minimum number of sub-bands directly from the
connectivity graph ${\cal G}$.

\section{Spectrum Division}\label{sec:MinColor}

In this section, we investigate the minimum number of sub-bands which yields a feasible
spectrum allocation. Our analysis will be based entirely on the network connectivity
graph ${\cal G}$. For convenience of exposition, we first transform the problem into an
equivalent graph-theoretic link-coloring problem.

\subsection{Link-Coloring Problem}

For the moment, we leave the total number of sub-bands undetermined. Let each sub-band be
identified by a unique color. We represent a spectrum allocation by a color assignment to
links. A feasible spectrum allocation corresponds to a color assignment such that (i) all
links are assigned with at least one color and (ii) for any node, no incoming link has a
common color with any one of its outgoing links. We will find the minimum number of
colors required for such a coloring. For any color assignment, denote the set of colors
used by node $i$'s outgoing links by ${\cal OC}_i$, and the set of colors used by its
incoming links by ${\cal IC}_i$. Since the network is connected and link-symmetric, every
node has at least one outgoing and one incoming link, implying that ${\cal OC}_i$ and
${\cal IC}_i$ are both nonempty for all $i \in {\cal N}$.  A color assignment is {\em
feasible} if and only if ${\cal OC}_i \cap {\cal IC}_i = \emptyset$ for all $i \in {\cal
N}$.\footnote{Note that we allow one link to transmit on multiple sub-bands. Hence, we do
not require $|{\cal OC}_i| \le |{\cal O}_i|$ or $|{\cal IC}_i| \le |{\cal O}_i|$.} Note
that the feasibility of a coloring scheme can be checked using the node color sets $\{
{\cal OC}_i, {\cal IC}_i\}_{i \in {\cal N}}$, regardless of the detailed assignment of
colors to links.  In fact, we will show that studying the node outgoing color sets $\{
{\cal OC}_i\}$ alone suffices for the link coloring problem. In the following, we say a
color assignment is {\em consistent} with $\{ {\cal OC}_i\}$ if $\{ {\cal OC}_i\}$
results from the color assignment. 

\vspace{0.1in}\begin{lemma}\label{lma:GenGraph1} Given a graph ${\cal G} = \{{\cal N},
{\cal L}\}$ and the nodes' outgoing color sets $\{ {\cal OC}_i\}_{i \in {\cal N}}$, there
exists a consistent and feasible color assignment if and only if

(i) ${\cal OC}_i \backslash {\cal OC}_j \ne \emptyset$, for all $(i,j) \in {\cal L}$;

(ii) $\bigcup_{j \in {\cal O}_i} {\cal OC}_i \backslash {\cal OC}_j = {\cal OC}_i$, for all $i \in {\cal N}$.
\end{lemma}\vspace{0.1in}

\textit{Proof:} We first prove the necessity part.  Suppose first that (i) is violated,
i.e., there exists a link $(i,j)$ such that ${\cal OC}_i \subseteq {\cal OC}_j$, then
$(i,j)$ must have a color in ${\cal OC}_j$, meaning that some outgoing link of $j$ has
the same color as $(i,j)$, which is infeasible. Next suppose that (ii) does not hold for
some node $i$, i.e., there exists a color $q \in {\cal OC}_i$ such that $q \in {\cal
OC}_j$ of every $j \in {\cal OC}_i$. Since $q$ is used on some outgoing link of $i$, say
$(i,k)$, $(i,k)$ will have a color belonging to ${\cal OC}_k$, violating the duplexing
constraint at node $k$.

The conditions are sufficient because given $\{{\cal OC}_i\}_{i \in {\cal N}}$ which
satisfy (i)-(ii), a feasible coloring can be constructed by assigning each link $(i,j)$
the color(s) in ${\cal OC}_i \backslash {\cal OC}_j$. By (ii), this coloring is also
consistent with $\{{\cal OC}_i\}_{i \in {\cal N}}$. \qed

In the following, we use the shorthand notation ${\cal OC}_i \nsim {\cal OC}_j$ to
represent ${\cal OC}_i \backslash {\cal OC}_j \ne \emptyset$ and ${\cal OC}_j \backslash
{\cal OC}_i \ne \emptyset$. A collection of node outgoing color sets $\{ {\cal OC}_i\}_{i
\in {\cal N}}$ is said to be {\em feasible} if it satisfies the conditions in
Lemma~\ref{lma:GenGraph1}.

We now state the problem of finding the minimum number of sub-bands in terms of the
minimal coloring problem as follows. Given a graph ${\cal G}=({\cal N}, {\cal L})$, \bea
\textrm{minimize} && \left| \bigcup_{i \in
{\cal N}}  {\cal OC}_i \right| \label{eq:GenMinColor} \\
\textrm{subject to} && \{ {\cal OC}_i\}_{i \in {\cal N}}~\textrm{feasible}. \nonumber
\eea Denote the minimum number of colors by $Q_{{\cal G}}$. To solve for $Q_{{\cal G}}$,
we first relax the constraints of problem~\eqref{eq:GenMinColor} by dropping the second
condition in Lemma~\ref{lma:GenGraph1} and consider \bea \textrm{minimize} && \left|
\bigcup_{i \in
{\cal N}}  B_i \right| \label{eq:GenMinColor1} \\
\textrm{subject to} &&  B_i \nsim B_j, ~~\forall  (i,j) \in {\cal L}, \nonumber \eea
where $B_i,~i \in {\cal N}$ are any non-empty sets. The optimal solution $Q_{\cal G}^*$
to \eqref{eq:GenMinColor1} must be less than or equal to $Q_{\cal G}$. We will find a
lower bound on $Q_{\cal G}^*$ and a matching upper bound on $Q_{\cal G}$, thus uniquely
determining $Q_{\cal G}$. For lower bounding $Q_{\cal G}^*$, the next observation is
useful.  The proof is deferred to Appendix~\ref{app:ProofLma2}.
\vspace{0.1in}\begin{lemma}\label{lma:CmpltGraph2} Let $B_1, \cdots, B_k$ be distinct
subsets of a $Q$-set all of which have the same cardinality $g$. If $g > \lfloor Q/2
\rfloor$, then there exist distinct subsets $C_1, \cdots, C_k$ such that $C_j \subset
B_j$ and $|C_j| = g-1$, $j = 1, \cdots, k$. If $g < \lfloor Q/2 \rfloor$, then there
exist distinct subsets $C_1, \cdots, C_k$ such that $B_j \subset C_j$ and $|C_j| = g+1$,
$j = 1, \cdots, k$.
\end{lemma}\vspace{0.1in}

Using Lemma~\ref{lma:CmpltGraph2} and the fact that any two distinct subsets $B_i$, $B_j$
with the same cardinality satisfy $B_i \nsim B_j$, we can show the following (see the
proof in Appendix~\ref{app:ProofLma3}).

\vspace{0.1in}\begin{lemma}\label{lma:CmpltGraph3} There exists an optimal solution
$\{B_i\}_{i \in {\cal N}}$ for~\eqref{eq:GenMinColor1} such
that all $B_i$, $i \in {\cal N}$, have the same cardinality. 
\end{lemma}\vspace{0.1in}

Using Lemma~\ref{lma:CmpltGraph3}, we can find a lower bound on $Q_{\cal G}^*$ as
follows. \vspace{0.1in}\begin{lemma}\label{lma:LowerBound} Given a graph ${\cal G} =
({\cal N}, {\cal L})$ with chromatic number $\chi({\cal G})$, the solution $Q_{\cal G}^*$
of the problem~\eqref{eq:GenMinColor1} is greater than or equal to $Q(\chi({\cal G}))$,
where the function $Q:\mathbb{Z}_+ \mapsto \mathbb{Z}_+$ is defined as
\be\label{eq:Qfunction} Q(N) \triangleq \min \left\{ q \in \mathbb{Z}_+:~ {{q} \choose
{\lfloor{q/2}\rfloor}} \ge N \right\}. \ee
\end{lemma}\vspace{0.1in}

\textit{Proof}: By Lemma~\ref{lma:CmpltGraph3}, we can without loss of optimality
consider a solution $\{B_i\}_{i \in {\cal N}}$ of~\eqref{eq:GenMinColor1} such that
$|B_i| = |B_j|$, for all $i,j \in {\cal N}$. Therefore, $B_i \nsim B_j$ whenever $B_i \ne
B_j$. Suppose there are $K$ distinct elements in $\{B_i\}_{i \in {\cal N}}$. Because
$|\bigcup_i B_i| = Q_{\cal G}^*$, by Sperner's theorem~\cite{paper:Spe28}, $K \le
{Q_{\cal G}^* \choose \lfloor Q_{\cal G}^* / 2 \rfloor}$. Moreover, since $\{B_i\}$
satisfies the constraint in~\eqref{eq:GenMinColor1}, we must have $K \ge \chi({\cal G})$,
where $\chi({\cal G})$ is the chromatic number of ${\cal G}$.  Therefore, $Q_{\cal G}^*
\ge Q(\chi({\cal G}))$. \qed

Lemma~\ref{lma:LowerBound} indicates that at least $Q(\chi({\cal G}))$ colors are needed
to construct a feasible collection of node color sets even with the condition (ii) in
Lemma~\ref{lma:GenGraph1} removed. However, by assigning node color sets according to a
minimal node-labelling scheme of ${\cal G}$, $Q(\chi({\cal G}))$ colors are sufficient to
yield a feasible configuration satisfying both conditions in Lemma~\ref{lma:GenGraph1}.
We can therefore conclude the following.

\vspace{0.1in}\begin{lemma}\label{lma:UpperBound} Given a graph ${\cal G} = ({\cal N},
{\cal L})$ with chromatic number $\chi({\cal G})$, the solution $Q_{\cal G}$ to the
problem~\eqref{eq:GenMinColor} is less than or equal to $Q(\chi({\cal G}))$.
\end{lemma}\vspace{0.1in}

\textit{Proof:} For graph ${\cal G}$, $Q(\chi({\cal G}))$ colors are sufficient because
they can generate at least $\chi({\cal G})$ color subsets $\{B_i\}$ with equal
cardinality $\lfloor Q(\chi({\cal G}))  / 2 \rfloor$, any two of which satisfy $B_i \nsim
B_j$. Associate any $\chi({\cal G})$ of the color subsets with $\chi({\cal G})$ labels.
Assign the $\chi({\cal G})$ color subsets to nodes according to a minimal node-labelling
scheme, which requires exactly $\chi({\cal G})$ labels. We then have a feasible
collection of node outgoing color sets that satisfies the two conditions in
Lemma~\ref{lma:GenGraph1}.\footnote{An arbitrary color subset assignment according to a
node-labelling solution generally results in $\{ {\cal OC}_i\}_{i \in {\cal N}}$
satisfying only (i) in Lemma~\ref{lma:GenGraph1}. Should (ii) not hold for some $i$,
reset ${\cal OC}_i := \bigcup_{j \in {\cal O}_i} {\cal OC}_i \backslash {\cal OC}_j$.
With this modification, both (i) and (ii) in Lemma~\ref{lma:GenGraph1} hold and no extra
color is needed.} \qed

%
%


We have from Lemmas~\ref{lma:LowerBound} and \ref{lma:UpperBound} that
\[
Q(\chi({\cal G})) \le Q_{{\cal G}}^* \le Q_{{\cal G}} \le Q(\chi({\cal G})).
\]
Therefore, all inequalities must hold with equality.

\vspace{0.1in}
\begin{proposition}\label{prop:General}
Let the chromatic number of graph ${\cal G}$ be $\chi({\cal G})$. Then, the solution of
\eqref{eq:GenMinColor} is $Q_{{\cal G}} = Q(\chi({\cal G}))$, where the function
$Q(\cdot)$ is given by~\eqref{eq:Qfunction}.
\end{proposition} \vspace{0.1in}

\begin{figure}[h]
\begin{center}
\includegraphics[width = 6cm]{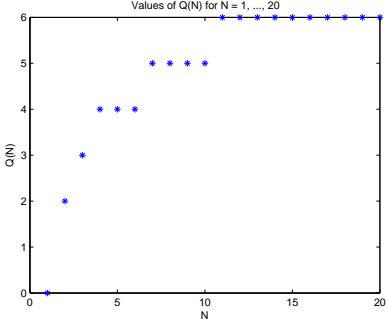}
\caption{The values of $Q(N)$, $N = 1, \cdots, 20$.}\label{fig:QFunction}
\end{center}
\end{figure}

It is easy to see that $Q(\cdot)$ is nondecreasing. The values of $Q(N)$ for $N = 1,
\cdots, 20$ are plotted in Figure~\ref{fig:QFunction}.
For large $N$, it follows from Stirling's approximation~\cite{book:Fel68}
\[
{{q} \choose {\lfloor{q/2}\rfloor}} = \sqrt{\frac{2}{\pi}}\exp \left[q \ln 2 -
\frac{1}{2} \ln q - \frac{1}{4q} + o\left(\frac{1}{q^2}\right)\right]
\]
that \[Q(N) = \Theta(\log N).\] Therefore by Proposition~\ref{prop:General}, $Q_{\cal G}$ grows at a \emph{logarithmic} rate with $\chi({\cal G})$. 


%
%
%

\section{Distributed Sub-band Allocation}\label{sec:Subband}

\subsection{Spectrum Division Using Approximation of $\chi({\cal G})$}

We have found that the minimum number of sub-bands depends on the chromatic number
$\chi({\cal G})$.  The problem of finding $\chi({\cal G})$,  however, is
NP-complete~\cite{book:GJ79} in general. On the other hand, any upper bound on
$\chi({\cal G})$ gives us a sufficient number of sub-bands.  One such well-known upper
bound is $\chi({\cal G}) \leq \Delta({\cal G})+1$, where $\Delta({\cal G})$ is the
maximum degree of any node in ${\cal G}$. In the following, we assume that there are at
least $Q(\Delta({\cal G})+1)$ sub-bands available.

The maximum degree $\Delta({\cal G})$ is straightforward to determine (in a distributed
manner), and we assume that it is known to all nodes a priori. Notice that
$Q(\Delta({\cal G})+1)$ is not too far from $Q(\chi({\cal G}))$ in typical
networks,\footnote{There are graphs, e.g. complete graphs, for which $\chi ({\cal G}) \le
\Delta({\cal G}) + 1$ holds with equality.} since
the $Q(\cdot)$ function is piecewise flat and grows approximately as a log function. 
It is worth comparing the upper bound $Q(\Delta({\cal G})+1)$ with $\Delta(\tilde{\cal
G})+1$, an upper bound of $\chi(\tilde{\cal G})$, the chromatic number of the
interference graph. Recall that vertices of $\tilde{\cal G}$ are (directed) links of
${\cal G}$. Any vertex $(i,j)$ of $\tilde{\cal G}$ has neighbors $\{(k,l)\in {\cal L}: k
= j~\textrm{or}~l = i, ~\textrm{but not both}\}$. Hence, by the assumption that ${\cal
G}$ is link-symmetric, the degree of $(i,j)$ in $\tilde{\cal G}$ is given by
\[
\Delta_{(i,j)} = |{\cal I}_i| + |{\cal O}_j|-1 = \Delta_i + \Delta_j - 1.\] Here,
$\Delta_i$ and $\Delta_j$ are the degrees of nodes $i$ and $j$ in ${\cal G}$.
Consequently, the maximum degree of $\tilde{\cal G}$ is
\[
\Delta(\tilde{\cal G}) = \max_{(i,j) \in {\cal L}} \Delta_i + \Delta_j - 1.
\]
On the other hand, $\Delta({\cal G}) = \max_{i \in {\cal N}} \Delta_i$. Therefore, it is
always true that
\[
\Delta(\tilde{\cal G}) \ge \Delta({\cal G}).
\]
The equality holds only when the network topology is extremely asymmetric, e.g. the star
networks. Typically, $\Delta(\tilde{\cal G})$ can be almost twice as large as
$\Delta({\cal G})$. Moreover, the $Q(\cdot)$ function is piecewise flat and scales as
$\log_2(\cdot)$. The difference of the two upper bounds $\Delta(\tilde{\cal G})+1 -
Q(\Delta({\cal G})+1)$ typically is substantial. Therefore, the connectivity graph
approach developed in this work not only is more simple and straightforward but also
leads to a
tighter upper bound than the conventional interference graph approach. 


\subsection{Distributed Sub-band Allocation Algorithm}

With the spectrum division accomplished, we now devise an algorithm with which feasible
$\{{\cal OC}_i\}$ can be determined in a distributed manner and a feasible spectrum
allocation can be found. The algorithm applies to all connected and link-symmetric
graphs. Assume that $Q$ sub-bands are available. For expositional purposes, we keep an
{\em unprocessed node set} ${\cal U}$, which initially contains all the nodes. The
Distributed Sub-band allocation (DSA) algorithm is iterated as follows.

\vspace{0.1in}{\bf Step 1.} Initially ${\cal U} = {\cal N}$. Arbitrarily select a node
$i$ from ${\cal U}$ and set ${\cal U} := {\cal U} \backslash \{i\}$. Arbitrarily choose
$\lfloor Q/2 \rfloor$ sub-bands to form ${\cal OC}_i$, mark $i$ as ``processed'', and go
to Step 2.

{\bf Step 2.}  Select an arbitrary node $i$ from ${\cal U}$ such that $i$ has at least
one processed neighbor (under the assumption that the network is connected, there is
always such a node in ${\cal U}$ after the first node is processed) and set ${\cal U} :=
{\cal U} \backslash \{i\}$. Node $i$ finds an ${\cal OC}_i$ with $\lfloor Q/2 \rfloor$
sub-bands which is different from all ${\cal OC}_j$ of its processed neighbors $j$ (it
can be shown that such an ${\cal OC}_i$ always exists). Moreover, to maximally avoid
potential channel interference, $i$ selects sub-bands in increasing order of their number
of occurrences\footnote{The number of occurrences can be zero.} in $\{{\cal OC}_j\}_{j
\in {\cal O}_i \backslash {\cal U}}$ (ties are broken arbitrarily).\footnote{For
instance, if the whole set of sub-bands is $\{R,G,B,Y\}$, and node $i$ finds its two
neighbors have chosen $\{R,G\}$ and $\{R,Y\}$, it will accordingly choose either
$\{B,G\}$ or $\{B,Y\}$.} Mark $i$ as ``processed''. If ${\cal U} \ne \emptyset$, repeat
Step 2; otherwise go to Step 3.

{\bf Step 3.}  Each node $i$ allocates the sub-band(s) in ${\cal Q}_{ij} := {\cal OC}_i
\backslash {\cal OC}_j$ to each outgoing link $(i,j)$. The algorithm terminates.

\vspace{0.1in}\begin{proposition}\label{prop:Subband1} The set $\{{\cal
Q}_{ij}\}_{(i,j)\in{\cal L}}$ generated by the DSA algorithm induces a feasible spectrum
allocation $\{{\cal L}_q\}_{q \in {\cal Q}}$.
\end{proposition}\vspace{0.1in}
\textit{Proof:} First we show that in Step 2 of the algorithm, an appropriate ${\cal OC}_i$ can always be found.
Because the whole set has $Q \ge Q(\Delta({\cal G})+1)$ elements, there are at least $\Delta({\cal G})+1$ distinct
subsets with cardinality $\lfloor Q/2 \rfloor$. However, $i$ has at most $\Delta({\cal G})$ (processed) neighbors.
Hence, there always exists at least one candidate for ${\cal OC}_i$.

Now we show that the $\{{\cal OC}_i\}$ obtained up to Step 3 are feasible. They satisfy
(i) in Lemma~\ref{lma:GenGraph1} due to the rule of successively selecting ${\cal OC}_i$
in Step 2. They also satisfy (ii) in Lemma~\ref{lma:GenGraph1}, as proved next. Suppose
on the contrary that there exists $i \in {\cal N}$ and $q \in {\cal OC}_i$ such that $q
\in {\cal OC}_j$ for all $j \in {\cal O}_i$. If $i$ is not the first processed node, the
fact that $q \in {\cal OC}_i$ implies that by the time ${\cal OC}_i$ is being determined,
each of the $Q - \lfloor Q/2 \rfloor$ sub-bands not included in ${\cal OC}_i$ must appear
in every previously processed neighbor $j$'s ${\cal OC}_j$. Taking also $q$ into account,
we can deduce that all those ${\cal OC}_j$ have at least $\lfloor Q/2 \rfloor + 1$
elements, which is a contradiction. Even if $i$ is the first processed node, the next
processed node $j$ must be $i$'s neighbor and ${\cal OC}_j \cup {\cal OC}_i = \emptyset$
due to the algorithm's rule. Hence, the hypothesis cannot be true in either case.

With the feasibility of the $\{{\cal OC}_i\}$ established, it is easy to verify that the
$\{{\cal Q}_{ij}\}_{(i,j)\in{\cal L}}$ generated in Step 3 induces a feasible spectrum
allocation $\{{\cal L}_q\}_{q \in {\cal Q}}$.  \qed

In practice, the DSA algorithm can be implemented in a distributed fashion by nodes in
the network. Specifically, after any node $i$ has arbitrarily set its ${\cal OC}_i$ at
the beginning, any other node $j$ can determine its own ${\cal OC}_j$ as long as it has
at least one processed neighbor and no other neighbor is being processed at the same
instant. In other words, it is possible to have two non-adjacent nodes configuring their
outgoing sub-bands at the same time. Since the action of a node depends only on its
neighbors, node operations need not be globally coordinated or synchronized across the
network.  All that is required is an initialization phase that designates the first node
to be processed. 
Thus, the sub-band allocation algorithm is distributed
and asynchronous in nature.

Finally, we note that the DSA algorithm is robust to dynamic node failure and addition.
Suppose that after a feasible spectrum allocation is established by the DSA algorithm, a
node leaves the network. In this case, the configuration for the remaining nodes and
links is still feasible. If a new node $i$ is to join the network, it can run Step 2 of
the DSA algorithm to determine its own ${\cal OC}_i$ based on its neighbors' ${\cal
OC}_j,j\in{\cal O}_i$. Then, active sub-bands for each link incident to $i$ can be
allocated as in Step 3. As long as $i$ connects to at most $\Delta({\cal G})$ nodes in
the old network ${\cal G}$, the algorithm yields a feasible spectrum allocation for node
$i$, with all other nodes' allocation unaffected.


\section{Cross-layer Optimization in Multi-radio Multi-hop Wireless Networks}\label{sec:Optimization}

The algorithm in Section~\ref{sec:Subband} generates a spectrum allocation where the
active links on every sub-band satisfy the duplexing constraints.  Given this feasible
spectrum allocation, we now develop corresponding joint power control, routing, and
congestion control algorithms which minimize total network cost, which will be specified
later, given channel conditions and traffic demands.

%

\subsection{Interference Limited Transmissions and Node Power Constraints}

We consider networks where messages on each sub-band are coded independently, and where
the receiver of a link decodes its message on an active sub-band while treating all other
signals on the same sub-band as interference. We assume that
the capacity $C_{ij}(q)$ of link $(i,j)$ on a sub-band $q$ is a function of $x_{ij}(q)$,
the signal-to-interference-plus-noise ratio (SINR) of link $(i,j)$ over $q$. Denoting the
transmission powers used by the active links on sub-band $q$ by
$\{P_{mn}(q)\}_{(m,n)\in{\cal L}_q}$, we have \be\label{eq:SINR} x_{ij}(q) =
\frac{G_{ij}^q P_{ij}(q)}{\sum_{\substack{(m,n) \in {\cal L}_q \\ (m,n) \ne (i,j)}}
G_{mj}^q P_{mn}(q) + N_j^q}, \ee where $G_{mj}^q$ is the path gain from $m$ to $j$ on
sub-band $q$, and $N_j^q$ is the power of the additive noise on sub-band $q$ at $j$. Note
that since the parameters $\{G_{ij}^q\}$ and $\{N_j^q\}$ are sub-band-dependent, this
framework is particularly appropriate for networks with frequency selective channels and
colored noise.

%

Assume each node is limited by an individual power constraint $\bar P_i$, i.e.,
\be\label{eq:PowerConstraint} \sum_{j\in{\cal O}_i} \sum_{q \in {\cal Q}_{ij}} P_{ij}(q)
\le \bar P_i. \ee Denote the set of power variables $\{P_{ij}(q)\}_{ (i,j) \in {\cal L},
q \in {\cal Q}_{ij}}$ that satisfy~\eqref{eq:PowerConstraint} by ${\cal P}$. In
Section~\ref{sec:Algorithms}, we will design a set of power control algorithms that
adjusts the power variables within the feasible region to minimize total network cost in
conjunction with congestion control and routing.

\subsection{Traffic Demands, Congestion Control, and Routing}

Let the traffic demands for the network consist of a collection ${\cal W}$ of unicast
sessions. Each (elastic) session $w \in {\cal W}$ is characterized by its fixed
source-destination node pair $(O(w), D(w))$ and demand rate $\bar r_w$.\footnote{We
assume that user $w$ gains no extra utility by transmitting at a rate higher than $\bar
r_w$. So without loss of optimality, $\bar r_w$ can be taken as the effective maximal
incoming rate demanded by $w$.} We model the traffic as fluid flows. Assume {\em
congestion control} is exercised at each source node. That is the source node $O(w)$ can
control the rate at which $w$'s traffic comes into the network. Denote the actual
admitted rate of $w$ by $r_w$, also referred to as the end-to-end flow rate of $w$. Thus,
the rate of the rejected traffic of $w$, denoted by $F_w$, is $\bar r_w - r_w$. After the
admitted flow rates are determined by the congestion control at the source nodes, the
flows entering the network are routed on (potentially) multiple paths from the source to
destination. Let $f_{ij}(w)$ denote the rate of session $w$ traffic routed through link
$(i,j)$. The session flows satisfy the {\em flow conservation} constraints, i.e.,
\be\label{eq:SourceFlow} \sum_{j \in {\cal O}_i} f_{ij}(w) = r_w, \quad i = O(w), \ee
\be\label{eq:InterFlow} \sum_{j \in {\cal O}_i} f_{ij}(w) = \sum_{k \in {\cal O}_i}
f_{ki}(w) =: t_i(w), \quad i \ne O(w),D(w), \ee and \be\label{eq:SinkFlow} f_{ij}(w) = 0,
\quad \forall j \in {\cal O}_i, \quad i = D(w).\ee Here, $t_i(w)$ denotes both the total
incoming and outgoing flow rates of session $w$ at an intermediate node $i$. If we
consider the rejected flow $F_w$ as being routed on a virtual ``overflow'' link from
$O(w)$ to $D(w)$~\cite{book:BG92,paper:XY06ISIT2}, the flow conservation constraint
involving both real and virtual outgoing flows from the source node is given by
\be\label{eq:VirtualFlow} \sum_{j \in {\cal O}_i} f_{ij}(w) + F_w = \bar r_w. \ee Thus,
later on we will incorporate congestion control in a unified framework with routing. Let
the set of flow vectors $(F_w, (f_{ij}(w))_{(i,j)\in{\cal L}})$ satisfying
\eqref{eq:InterFlow}-\eqref{eq:VirtualFlow} be denoted by ${\cal F}_w$.

The total flow rate on link $(i,j)$ is $F_{ij} = \sum_{w \in {\cal W}}f_{ij}(w)$. To
route a flow of rate $F_{ij}$ from $i$ to $j$, node $i$ can split the traffic onto all
active sub-bands of link $(i,j)$ and transmit them simultaneously. Let the rate of flow
assigned to sub-band $q \in {\cal Q}_{ij}$ be $F_{ij}(q)$. We have $F_{ij} = \sum_{q \in
{\cal Q}_{ij}} F_{ij}(q)$.  Hence, \be\label{eq:FlowEquation} \sum_{q \in {\cal Q}_{ij}}
F_{ij}(q) = \sum_{w \in {\cal W}}f_{ij}(w). \ee Note that the flow on each sub-band may
consist of traffic of one or more sessions. However, the traffic split conserves the
total rate of any session going through link $(i,j)$. Therefore, at the receiver end of
the link, node $j$ collects each session $w$'s  traffic of rate $f_{ij}(w)$.

One can think of the routing scheme as a two-step process. The first step determines {\em
inter-node routing}, i.e., a feasible flow vector $(F_w, (f_{ij}(w))_{(i,j)\in{\cal L}})
\in {\cal F}_w$ is found for every $w \in {\cal W}$, yielding $(F_{ij})_{(i,j)\in{\cal
L}}$. The second step amounts to {\em intra-node routing}: each node $i$ routes the flow
$F_{ij}$ onto the active sub-bands $q \in {\cal Q}_{ij}$ available from $i$ to $j$.


\subsection{Network Cost and Optimal Resource Allocation}

The cost on each (real) link $(i,j)$ is made up of the costs $D_{ij}^q$ incurred on its
active sub-band(s) $q \in {\cal Q}_{ij}$. Each $D_{ij}^q$ is a function of $x_{ij}(q)$,
the SINR of link $(i,j)$ on $q$, and $F_{ij}(q)$, the flow rate on the active sub-band
$q$ of link $(i,j)$. Specifically, define
\[ D_{ij}^q = D(x_{ij}(q), F_{ij}(q)).
\]
Assume that $D(x, F)$ is decreasing and convex in $x$ for fixed $F$, and that it is
increasing and convex in $F$ for fixed $x$. To impose the hard capacity constraint $F <
C(x)$, where $C(x)$ denotes the link capacity achieved by having $SINR = x$, one can
define $D(x, F) = \infty$ for $F \ge C(x)$.  An example of such cost functions
is\footnote{For modelling purposes, we assume that on the transmitter side of each link
there is one queue for every active sub-band of that link. Specifically, the arrival rate
to the queue of link $(i,j)$ on sub-band $q$ is $F_{ij}(q)$, the aggregate flow routed
onto $(i,j)$ through $q$. The cost function $D(x,F) = \frac{F}{C(x)-F}$ gives the
expected delay in an M/M/1 queue with arrival rate $F$ and service capacity $C(x)$. In
our present setting, $F$ is the sub-band specific link flow rate and $C(x)$ is the
sub-band specific link capacity. By the Kleinrock independence approximation and
Jackson's Theorem, the $M/M/1$ queue is a good approximation for the behavior of
individual links on active sub-bands when the system involves Poisson stream arrivals at
the entry points, a densely connected network, and moderate-to-heavy traffic load
\cite{book:Kle64,book:BG92}.} $D(x,F) = \frac{F}{C(x)-F}$ where $C(x)=R\log(Kx)$ with
constants $R, K
> 0$.\footnote{The constants $R,K$
correspond to the bandwidth and processing gain of the system. The capacity formula
$C(x)=R\log(Kx)$ is a good approximation for systems with a high processing gain, e.g.
CDMA networks.}

We associate an increasing and concave utility function $U_w(r_w)$, $0 \le r_w \le \bar
r_w$, to each (elastic) session $w$. Thus, rejecting a flow of $F_w$ from the network
incurs a utility loss of $U_w(\bar r_w) - U_w(\bar r_w - F_w)$. This utility loss can
also be interpreted as the cost $D_w(F_w) \triangleq U_w(\bar r_w) - U_w(\bar r_w - F_w)$
on a virtual overflow link~\cite{book:BG92} with flow rate $F_w$. Note that by
definition, $D_w(F_w)$ is increasing and convex in $F_w$.

We solve for the feasible flow and power variables that jointly maximize the net utility,
i.e., total utility minus total cost on real links. With the above observation, the
maximization of net utility is equivalent to the problem of minimizing the total network
cost $E \triangleq \sum_{(i,j) \in {\cal L}}D_{ij} + \sum_{w \in {\cal W}} D_w$
consisting of costs on both real and overflow links. Formally, the Multi-Radio Minimum
Cost Resource Allocation (MCRA) problem is \bea \textrm{minimize} && \sum_{w\in{\cal W}}
D_w(F_w)+\sum_{(i,j) \in
{\cal L}}\sum_{q \in {\cal Q}_{ij}} D(x_{ij}(q), F_{ij}(q))  \nonumber \\
\textrm{subject to} &&\eqref{eq:SINR}~\textrm{and}~(P_{ij}(q))_{ (i,j) \in {\cal L}, q
\in {\cal Q}_{ij}} \in {\cal P}, \nonumber \\
&&\eqref{eq:FlowEquation}~\textrm{and}~(F_w,
(f_{ij}(w))_{(i,j)\in{\cal L}}) \in {\cal F}_w, \; \forall w
\in{\cal W}.\nonumber\eea

We will devise a distributed scheme that iteratively adjusts power control, routing and
congestion control on a node-by-node basis so as to find the optimal solution to the MCRA
problem.

\subsection{Control Variables and Optimality Conditions}

To update transmission powers, traffic routes, and traffic input rates, we define the
following power control and routing variables. These optimization variables are
node-based and local in the sense that they are independently controlled at individual
nodes.

\subsubsection{Power Control Variables}\label{subsec:PowerVariables}

Let $P_i(q)$ be the total power used by node $i$ on its active sub-band $q$. Let
$P_{ij}(q)$ be the power used by node $i$ on its outgoing link $(i,j)$ over an active
sub-band $q$.  Define
\[
\rho_i(q) \triangleq \frac{P_i(q)}{\bar P_i}, \quad q \in {\cal OC}_i,\] constrained by
\be\label{eq:RhoConstraint} \quad 0 \le \rho_i(q) \le 1, \quad \sum_{q \in {\cal OC}_i}
\rho_i(q) \le 1. \ee Also define
\[
\eta_{ij}(q) \triangleq \frac{P_{ij}(q)}{P_i(q)}, \quad j\in {\cal O}_i, q \in {\cal
Q}_{ij},\] subject to constraints (let $\eta_{ij}(q) \equiv 0$ for $q \notin {\cal
Q}_{ij}$) \be\label{eq:EtaConstraint} \quad 0 \le \eta_{ij}(q) \le 1, \quad \sum_{j \in
{\cal O}_i} \eta_{ij}(q) = 1. \ee

With the above definitions, $x_{ij}(q)$ can be expressed in terms of the power control
variables as
\begin{small}\be\label{eq:SINR'} \frac{G_{ij}^q \bar P_i \rho_i(q) \eta_{ij}(q)}{G_{ij}^q \bar P_i \rho_i(q)
\sum_{n \ne j} \eta_{in}(q) + \sum_{m \ne i}G_{mj}^q \bar P_m \rho_m(q) + N_j^q}. \ee
\end{small}We use $IN_{ij}(q)$ to denote the total interference and noise power of link $(i,j)$ on sub-band $q$, i.e. the denominator of \eqref{eq:SINR'}.

Define $\bs\rho_i \triangleq (\rho_i(q))_{q \in {\cal OC}_i}$ and $\bs\eta_i \triangleq
(\eta_{ij}(q))_{j\in {\cal O}_i, q \in {\cal Q}_{ij}}$. We now compute the partial
derivative of the network cost $E$ with respect to each $\rho_i(q)$ and $\eta_{ij}(q)$.
These partial derivatives are useful for characterizing the optimality conditions and for
iterative adjustment of the power control variables. We have \be\label{eq:DelEta}
\frac{\partial E}{\partial \eta_{ij}(q)} = P_i(q) \left\{ \sum_{n \in {\cal O}_i}
(D_{x_{in}}^q)' \frac{- G_{in}^q x_{in}(q)} { IN_{in}(q)} + \delta\eta_{ij}(q) \right\},
\ee where we have used shorthand notation $(D_{x_{in}}^q)'$ for $\partial D(x_{in}(q),
F_{in}(q)) /
\partial x_{in}(q)$ and defined \be\label{eq:DeltaEta} \delta\eta_{ij}(q) \triangleq
(D_{x_{ij}}^q)' \frac{ G_{ij}^q (1 + x_{ij}(q))}{ IN_{ij}(q)}.\ee Note that
$\delta\eta_{ij}(q)$ as well as ${\partial E}/{\partial \eta_{ij}(q)}$ involves only {\em
local measures} of $i$. The partial derivative of $E$ with respect to $\rho_i(q)$ is
given by
\begin{small}\be\label{eq:DeltaRho} \delta\rho_i(q) \triangleq \frac{\partial E}{\partial
\rho_{i}(q)} = \bar P_i \left\{ \sum_{n \in {\cal N}} G_{in}^q MSG_n^q + \sum_{j \in
{\cal O}_i} {\delta\eta_{ij}(q) \eta_{ij}(q)} \right\},\ee \end{small} where
\be\label{eq:MSG} MSG_n^q \triangleq \sum_{m \in {\cal O}_n} (D_{x_{mn}}^q)' \frac{-
G_{mn}^q P_{mn}(q)}{IN_{mn}(q)^2} \ee is the message that node $n$ needs to send to any
other node, say $i$, which is active on $q$ in order for $i$ to compute $\delta\rho_i(q)$
via \eqref{eq:DeltaRho}. Specifically, the message exchange works as follows.

\begin{figure}[h]
\begin{center}
\includegraphics[width = 7cm]{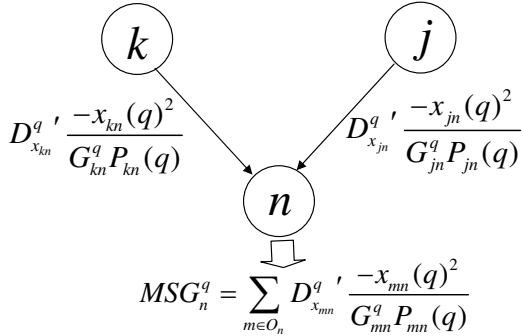}
\caption{Power control message generation.}\label{fig:PC}
\end{center}
\end{figure}

\emph{Power Control Message Exchange Protocol:} Let each node $n$ keep an up-to-date
measure $MSG_n^q$ for each sub-band $q$, where $MSG_n^q$ is derived by assembling the
measures \[(D_{x_{mn}}^q)' \frac{- G_{mn}^q P_{mn}(q)}{IN_{mn}(q)^2} = (D_{x_{mn}}^q)'
\frac{- x_{mn}(q)^2}{G_{mn}^q P_{mn}(q)}\] on all its active incoming links $(m,n)$ on
$q$, and summing them up (cf.~\eqref{eq:MSG}). Note that $MSG_n^q$ is nonnegative and it
is zero if $n$ has no active incoming link on $q$. If $MSG_n^q > 0$, node $n$ broadcasts
it to the whole network via a flooding protocol. This control message generating process
is illustrated by Figure \ref{fig:PC}.
If $q \in {\cal OC}_i$, node $i$ collects $MSG_n^q$ and processes it according to the
following rule. Node $i$ multiplies $MSG_n^q$ with the path gain $G_{in}^q$.\footnote{In
a symmetric duplex channel, $G_{in}^q \approx G_{ni}^q$, and node $i$ may use its own
measurement of $G_{ni}^q$ in the place of $G_{in}^q$. Otherwise, it will need channel
feedback from node $n$ to calculate $G_{in}^q$.} It further adds the product to the value
of local measure $\delta\eta_{in}(q) \cdot \eta_{in}(q)$ if the origin $n$ of the message
is a neighbor of $i$ and $(i,n)$ is active on sub-band $q$. Finally, node $i$ adds up all
the processed messages, and this sum multiplied by $\bar P_i$  equals $\delta\rho_i(q)$.
Note that this protocol requires {\em only one message from each node} on each sub-band
$q$.\footnote{To be more precise, only the nodes having at least one active incoming link
on a certain sub-band need to provide a message for that sub-band.} Moreover in practice,
node $i$ can ignore the messages generated by distant nodes, because they contribute very
little to $\delta\rho_i(q)$ due to the negligible multiplicative factor $G_{in}^q$ on
$MSG_n^q$ when $i$ and $n$ are far apart (cf.~\eqref{eq:DeltaRho}).

\subsubsection{Routing Variables}\label{subsec:RoutingVariables}

Routing variables were first introduced by Gallager \cite{paper:Gal77} for wireline
network routing problems. Here, we define routing variables in a similar fashion. In
addition to inter-node routing, however, routing variables here also perform the function
of congestion control and intra-node routing.

\begin{figure}[h]
  \begin{center}
  \includegraphics[width=7cm]{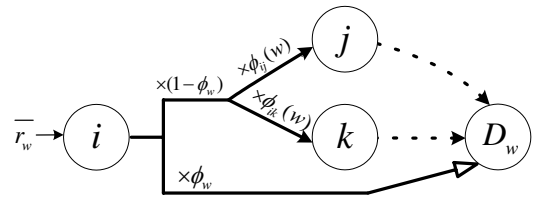}\\
  \caption{Overflow and inter-node routing at a source node.}\label{fig:OverflowRouting}
  \end{center}
\end{figure}

\begin{figure}[h]
  \begin{center}
  \includegraphics[width=7cm]{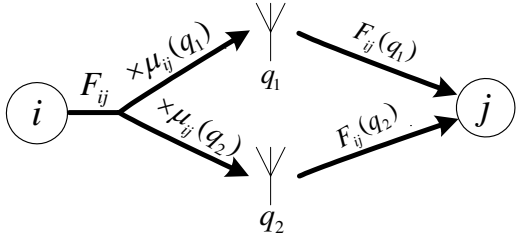}\\
  \caption{Intra-node routing on link $(i,j)$.}\label{fig:IntraRouting}
  \end{center}
\end{figure}

Recall that congestion control is equivalent to routing a portion of traffic demand on a
virtual overflow link directly from the source to the destination. Let $i$ be the source
node of session $w$, define overflow routing variable
\[
\phi_w \triangleq \frac{F_w}{\bar r_w},
\]
which is constrained by $0 \le \phi_w \le 1$.  The overflow rate is then controlled by
$\phi_w$ as $F_w = \bar r_w \phi_w$, and the end-to-end flow rate is given by $r_w = \bar
r_w (1 - \phi_w) $. Routing variables associated with a real link $(i,j)$ are defined by
\[
\phi_{ij}(w) \triangleq \frac{f_{ij}(w)}{t_i(w)},
\]
which gives the fraction of the incoming flow of session $w$ at
node $i$ that is routed onto link $(i,j)$. At any node $i$ except
the destination, $\phi_{ij}(w)$ of all $j \in {\cal O}_i$ satisfy
\be\label{eq:PhiConstraint} 0\le\phi_{ij}(w) \le 1
\quad\textrm{and}\quad \sum_{j \in {\cal O}_i} \phi_{ij}(w) = 1.
\ee It is easy to see that the routing variables $\phi_w,
\phi_{ij}(w)$ of all $(i,j)\in {\cal L}$ and $w \in {\cal W}$
uniquely determine the inter-node flow patterns $(F_w,
(f_{ij}(w))_{(i,j)\in{\cal L}})$ of all sessions $w$, and hence
the total flow rate on links $F_{ij} = \sum_w f_{ij}(w)$. Now let
the flow allocation on active sub-bands be specified by the
intra-node routing variables defined by
\[
\mu_{ij}(q) \triangleq \frac{F_{ij}(q)}{F_{ij}}, \quad q \in {\cal Q}_{ij}.
\]
For any link $(i,j)$, we must have \be\label{eq:MuConstraint} 0 \le \mu_{ij}(q) \le 1
\quad\textrm{and}\quad \sum_{q \in {\cal Q}_{ij}} \mu_{ij}(q) = 1. \ee The link flow rate
on an active sub-band is therefore given by
\[
F_{ij}(q) = \mu_{ij}(q)\sum_{w \in {\cal W}} t_i(w) \phi_{ij}(w), \quad q \in {\cal
Q}_{ij}.
\]
For notational brevity, denote by $\bs\phi_i(w)$ the vector $(\phi_{ij}(w))_{j\in{\cal
O}_i}$ if $i \neq O(w), D(w)$, or $(\phi_w, (\phi_{ij}(w))_{j\in{\cal O}_i})$ if $i =
O(w)$. Also denote by $\bs\mu_i$ the vector $(\mu_{ij}(q))_{j \in {\cal O}_i, q \in {\cal
Q}_{ij}}$. We illustrate the use of the above routing variables by looking at a source
node $i$ and one of its outgoing links $(i,j)$ in Figures \ref{fig:OverflowRouting} and
\ref{fig:IntraRouting}. In Figure \ref{fig:OverflowRouting}, the overflow link is marked
by a solid line with hollow arrow.
In Figure \ref{fig:IntraRouting}, we assume that link $(i,j)$ is
active on two sub-bands $q_1$ and $q_2$, and that the inter-node
flow rate from $i$ to $j$ is $F_{ij}$.

We now compute the partial derivative of $E$ with respect to those routing variables.
First we have the derivative of $E$ with respect to the overflow routing variables
\[
\frac{\partial E}{\partial \phi_w} = \bar r_w \left( D_w' - \frac{\partial E}{\partial
t_{w}} \right),
\]
where $\frac{\partial E}{\partial t_{w}}$ is the marginal cost of increasing the
end-to-end flow rate of session $w$ with all other variables held constant. This is a
special case ($i = O(w)$) of $\frac{\partial E}{\partial t_{i}(w)}$, which is the
marginal cost of increasing the incoming flow rate of $w$ at node $i$ while keeping all
other variables constant. These marginal costs are computed recursively as
in~\cite{paper:Gal77}:
\[
\frac{\partial E}{\partial t_{i}(w)} = 0, \quad \textrm{if~}i = D(w),
\]
and for $i \neq D(w)$, \begin{small}\bea \frac{\partial E}{\partial t_{i}(w)} &=&
\sum_{j\in{\cal O}_i} \phi_{ij}(w) \left( \sum_{q\in{\cal Q}_{ij}} \mu_{ij}(q)
{D_{F_{ij}}^q}' + \frac{\partial
E}{\partial t_{j}(w)} \right)\nonumber\\
&=& \sum_{j\in{\cal O}_i} \phi_{ij}(w) \delta\phi_{ij}(w), \label{eq:RoutingMargCost}
\eea \end{small}where we have used shorthand notation ${D_{F_{ij}}^q}'$ for $\partial
D(x_{ij}(q), F_{ij}(q)) /
\partial F_{ij}(q)$ and defined {\em marginal routing cost indicators}
\be\label{eq:DeltaPhi} \delta\phi_{ij}(w) = \sum_{q\in{\cal Q}_{ij}} \mu_{ij}(q)
{D_{F_{ij}}^q}' + \frac{\partial E}{\partial t_{j}(w)}. \ee Using $\delta\phi_{ij}(w)$,
we can easily write out the partial derivative of $E$ in $\phi_{ij}(w)$ as
\be\label{eq:DelPhi}
\frac{\partial E}{\partial \phi_{ij}(w)} = t_i(w)
\delta\phi_{ij}(w).
\ee
Finally, the partial derivative of $E$ in $\mu_{ij}(q)$ is given
by
\begin{equation}\label{eq:DelMu}
\frac{\partial E}{\partial \mu_{ij}(q)} = F_{ij} {D_{F_{ij}}^q}'.
\end{equation}
Notice that $\frac{\partial E}{\partial \mu_{ij}(q)}$ depends on {\em local measures}
only. However, $\frac{\partial E}{\partial \phi_w}$ and $\frac{\partial E}{\partial
\phi_{ij}(w)}$ are both tied to the value of $\frac{\partial E}{\partial t_{j}(w)}$ of
$i$'s downstream\footnote{Given a routing configuration $\{\phi_{ij}(w)\}_{(i,j)\in{\cal
L}}$ of session $w$, node $j$ is said to be {\em downstream} to $i$ if there exists a
path $(i,j_1)$, $(j_1,j_2)$, $\cdots$, $(j_n, j)$ such that $\phi_{ij_1}$,
$\phi_{j_1j_2}$, $\cdots$, $\phi_{j_nj}$ are all positive. We say $i$ is upstream to $j$
with respect to session $w$ if $j$ is downstream to $i$.} neighbors $j$, which in turn
depends on $\frac{\partial E}{\partial t_{k}(w)}$ of $j$'s downstream neighbors $k$.
Thus, we need a sequential message passing of the marginal routing costs, from the
destination upstream to the source, to permit every node to acquire the partial
derivatives in its local routing variables.

\emph{Routing Message Exchange Protocol:} In \cite{paper:Gal77}, the rules for
propagating the marginal routing cost information are specified. In order for node $i$ to
evaluate the terms $\delta\phi_{ij}(w)$ in \eqref{eq:DeltaPhi}, it needs to collect local
measures ${D_{F_{ij}}^q}'$ from all $q \in {\cal Q}_{ij}$ as well as reports of marginal
costs ${\partial D}/{\partial r_j(w)}$ from its next-hop neighbors $j \in {\cal O}_i$,
for all traversing sessions $w \in {\cal W}$. Moreover, it is responsible for calculating
its own measure of marginal cost $\frac{\partial D}{\partial t_i(w)}$ with respect to
every session $w$ according to \eqref{eq:RoutingMargCost}, and then providing the measure
to its upstream neighbors with respect to the session $w$. The sequential message passing
terminates if and only if the routing pattern of the session contains no loops, which is
guaranteed by the blocked-node-set technique developed in \cite{paper:Gal77,paper:BGG84}.

%

\subsubsection{Conditions for Optimality}

The power and routing configuration that solves the MCRA problem can be characterized in
terms of the marginal power and routing costs as follows.

\vspace{0.1in}\begin{theorem}\label{thm:OptCondition} For an
instance of optimization variables $\{ \bs \rho_i, \bs\eta_i,
\{\bs\phi_i(w)\}_{w\in{\cal W}}, \bs\mu_i \}_{i \in {\cal N}} $ to
be optimal, the following conditions are necessary: For all $i \in
{\cal N}$ and $w \in {\cal W}$ with $t_i(w) > 0$,
\be\label{eq:OptRouting1} \frac{\partial E}{\partial t_i(w)} =
\min_{j \in {\cal O}_i}\left\{ \min_{q \in {\cal Q}_{ij}} \left[
{D_{F_{ij}}^q}' \right] + \frac{\partial E}{\partial t_j(w)}
\right\}. \ee Moreover, for all $w \in {\cal W}$,
\be\label{eq:OptRouting2} \frac{\partial E}{\partial t_{w}}
\left\{\ba{ll}
                                            \le D_w', \quad &\textrm{if~}\phi_w = 0,\\
                                            = D_w', &\textrm{if~}0 < \phi_w < 1,\\
                                            \ge D_w', &\textrm{if~} \phi_w = 1.
                                            \ea \right.
\ee For all $i \in {\cal N}$ and $q \in {\cal OC}_i$, there exists a $\lambda_i$ such
that \be\label{eq:OptPower1} \delta\rho_i(q) \left\{\ba{ll} = \lambda_i,
\quad&\textrm{if}~ \rho_i(q) > 0,\\
\ge \lambda_i, &\textrm{if}~\rho_i(q) = 0, \ea\right. \ee and the constant $\lambda_i$
satisfies \be\label{eq:OptPower2} \lambda_i \left\{\ba{ll}
                                            \ge 0, \quad &\textrm{if~} \sum_{q\in{\cal
                                                                            OC}_i}\rho_i(q) =0,\\
                                                                            =0, &\textrm{if~} 0<\sum_{q\in{\cal
                                                                            OC}_i}\rho_i(q) < 1,\\
                                                                            \le 0, &\textrm{if~} \sum_{q\in{\cal
                                                                            OC}_i}\rho_i(q) =
                                                                            1.
                                                                            \ea \right. \ee
Furthermore, for $q \in {\cal OC}_i$ such that $\rho_i(q) > 0$,
there exists a constant $\gamma_i(q)$ such that for all $j \in
{\cal O}_i$ with $q \in {\cal Q}_{ij}$, \be\label{eq:OptPower3}
\delta\eta_{ij}(q) \left\{ \ba{ll}
                            = \gamma_i(q), \quad &\textrm{if}~\eta_{ij}(q)>0,\\
                            \ge \gamma_i(q), &\textrm{if}~\eta_{ij}(q)=0.
                            \ea
                            \right. \ee

\end{theorem}\vspace{0.1in}

We sketch the proof of Theorem~\ref{thm:OptCondition} in Appendix~\ref{app:ProofThm1}.
First order conditions as in \eqref{eq:OptRouting1}-\eqref{eq:OptPower3} are in general
only necessary for a configuration to be optimal. However, they are further sufficient if
the link cost function $D(x,F)$ has certain convexity properties as we discuss next.

Let $x$ generically represent the SINR of a link on one of its active sub-bands. By
\eqref{eq:SINR}, $x$ is a function of the power variables of all active links on the same
sub-band, e.g. $x_{ij}(q) = h_{ij}^q((P_{mn}(q))_{(m,n)\in{\cal L}_q})$. Thus, $D_{ij}^q$
is a function of the vector $\bs P(q) \triangleq (P_{mn}(q))_{(m,n)\in{\cal L}_q}$ and
$F_{ij}(q)$. It turns out that a characterization of the sufficient conditions for
optimality requires $D_{ij}^q$ to be jointly convex in $\bs P(q)$ and $F_{ij}(q)$.
However, it can be shown that such a property is impossible given the assumption that
$D_{ij}^q$ is decreasing in $x_{ij}(q)$.

One way to remedy this problem is to work with log-power variables, first introduced in
\cite{paper:JXB03,paper:Chi04}. For each $(m,n) \in {\cal L}$ and $q \in {\cal Q}_{mn}$,
define $S_{mn}(q) = \ln P_{mn}(q)$. Consider $D_{ij}^q$ as a function of $\bs S(q)
\triangleq (S_{mn}(q))_{(m,n)\in{\cal L}_q}$ and $F_{ij}(q)$. If $D_{ij}^q$ is jointly
convex in $\bs S(q)$ and $F_{ij}(q)$, the conditions in Theorem \ref{thm:OptCondition}
are enough to imply the optimality of a configuration. The joint convexity property holds
if and only if the link cost function $D(x,F)$ satisfies the following.

\vspace{0.1in}\begin{lemma}\label{lma:ConvexCost} For all $(i,j) \in {\cal L}$ and $q \in
{\cal Q}_{ij}$, $D_{ij}^q$ is jointly convex in $\bs S(q)$ and $F_{ij}(q)$ if and only if
the matrix \be\label{eq:Mmatrix} M = \left[ \ba{cc} D_x'' x^2 + D_x' x \; & D_{xF}'' x \\
D_{xF}'' x & D_F''\ea \right] \ee is positive semidefinite for all $(x,F)$.
\end{lemma}\vspace{0.1in}

In \eqref{eq:Mmatrix}, $D_x'$ and $D_x''$ denote the first and second partial derivatives
of $D(x,F)$ with respect to $x$, $D_F''$ denotes the second partial derivative of $D$
with respect to $F$, and $D_{xF}''$ denotes $\partial^2 D / \partial x \partial F$.
It can be shown
 that the cost function $D(x,F) = \frac{F}{C(x)-F}$ with $C(x)=R\log(Kx)$
satisfies the condition in Lemma~\ref{lma:ConvexCost}. Limited by the available space, we
skip the proof of Lemma~\ref{lma:ConvexCost}. We prove in Appendix~\ref{app:ProofThm2}
that if the condition in Lemma \ref{lma:ConvexCost} holds, the conditions stated in the
next theorem are sufficient for optimality.

\vspace{0.1in}\begin{theorem}\label{thm:SuffOptCond}

If the matrix $M$ in \eqref{eq:Mmatrix} of link cost function $D(x,F)$ is positive
semidefinite, then conditions \eqref{eq:OptRouting1}-\eqref{eq:OptPower3} are sufficient
for $\{ \bs \rho_i, \bs\eta_i, \{\bs\phi_i(w)\}_{w\in{\cal W}}, \bs\mu_i \}_{i \in {\cal
N}} $ to be optimal if \eqref{eq:OptRouting1} holds at every node $i \ne D(w)$ whether
$t_i(w) > 0$ or not, and if \eqref{eq:OptPower3} holds for all $i \in {\cal N}$ and $q
\in {\cal OC}_i$ whether $\rho_i(q) > 0$ or not.
\end{theorem}\vspace{0.1in}


\section{Node-Based Multi-Radio Power Control and Routing Algorithms}\label{sec:Algorithms}

We develop a set of scaled gradient projection algorithms~\cite{book:Ber99} by which
individual nodes adjust their local power control and routing variables iteratively to
achieve a global configuration satisfying the optimality conditions in
Theorem~\ref{thm:SuffOptCond}.  Each algorithm at a node updates the appropriate vector
of local optimization variables, e.g. $\bs\rho_i$, $\bs\eta_i$, $\bs\phi_i(w)$ or
$\bs\mu_i$, such that the updated vector results in a lower cost with all other variables
held constant. This is achieved by updating the old vector in the opposite gradient
direction scaled by an appropriate positive definite matrix, and projecting the new
vector back into the feasible set whenever it falls outside. This technique has been
applied to, for example, optimal routing in wireline networks~\cite{paper:BGG84} and
optimal power control and routing in single-radio wireless
networks~\cite{paper:XY06ISIT1}.

\subsection{Power Control Algorithms}

We develop two sets of power control algorithms which let each node $i$ iteratively
adjust the vector $(\eta_{ij}(q))_{j \in {\cal O}_i}$ and the vector $(\rho_i(q))_{q \in
{\cal OC}_i}$. First notice that if $q \in {\cal OC}_i$ but ${\cal L}_q$ contains only
one of $i$'s outgoing links, say $(i,j)$, $\eta_{ij}(q)$ must be equal to $1$, and hence
is not variable. Otherwise, if $\bs\eta_i(q) \triangleq (\eta_{ij}(q))_{(i,j) \in {\cal
L}_{q}}$ has more than one positive element, it is updated by the following scaled
gradient projection algorithm: \be\label{eq:EtaAlg} \bs\eta_i^{k+1}(q) := \left[
\bs\eta_i^{k}(q) - \left( Q_i^k(q) \right)^{-1} \cdot \delta\bs\eta_i^{k}(q)
\right]_{Q_i^k(q)}^+. \ee Here, the superscripts $k$, $k+1$ are the iteration indices,
$Q_i(q)$ is a positive definite scaling matrix, $\delta\bs\eta_i(q)$ is the vector
$(\delta\eta_{ij}(q))_{(i,j) \in {\cal L}_{q}}$, and $[\cdot]_{Q_i^k(q)}^+$ is the
projection operation onto the feasible set of $\bs\eta_i(q)$
(cf.~\eqref{eq:EtaConstraint}) relative to the norm induced by $Q_i^k(q)$.\footnote{In
general, $[\tilde{\bs x}]_{M}^+ \triangleq \arg\min_{\bs x \in {\cal F}} (\bs x -
\tilde{\bs x})' \cdot M \cdot (\bs x - \tilde{\bs x})$, where ${\cal F}$ is the feasible
set of $\bs x$.} To implement the algorithm~\eqref{eq:EtaAlg}, node $i$ needs the current
value of $\delta\bs\eta_i(q)$, which by \eqref{eq:DeltaEta} is easily computable from
local measures. 
%

\vspace{0.1in}The algorithm used by node $i$ to iteratively update $\bs\rho_i$ is as
follows: \be\label{eq:RhoAlg} \bs\rho_i^{k+1} := \left[ \bs\rho_i^{k} - \left( V_i^k
\right)^{-1} \cdot \delta\bs\rho_i^{k} \right]_{V_i^k}^+. \ee It has almost the same form
as~\eqref{eq:EtaAlg} except that the feasible set for the projection is defined by
\eqref{eq:RhoConstraint}. However, notice that each component of $\delta\bs\rho_i^{k}
\triangleq (\delta\rho_i(q)^k)_{q \in {\cal OC}_i}$ depends on measures from all active
links on sub-band $q$. Therefore, prior to implementing~\eqref{eq:RhoAlg}, node $i$ needs
to collect the appropriate power control messages to determine $\delta\bs\rho_i^{k}$, as
described by the message exchange protocol in Section~\ref{subsec:PowerVariables}.

\subsection{Routing Algorithms}

We now present the algorithm used by a node $i$ to update its inter-node routing vector
$\bs\phi_i(w)$ for a traversing session $w$. An iteration of the inter-node routing
algorithm has the form \be\label{eq:InterNodeRT} \boldsymbol\phi_i^{k + 1}(w) :=
\left[\boldsymbol\phi_i^k(w) - (M_i^k(w))^{-1} \cdot
\delta\boldsymbol\phi_i^k(w)\right]_{M_i^k(w)}^+.  \ee Here,
$\delta\boldsymbol\phi_i^k(w) \triangleq (\delta\phi_{ij}^k(w))_{j \in {\cal O}_i}$ is
the vector of current marginal routing cost indicators and $M_i^k(w)$ is a positive
definite matrix used to scale the descent direction. The feasible set associated with the
projection operation in \eqref{eq:InterNodeRT} is prescribed by the constraint
\eqref{eq:PhiConstraint} and an additional requirement that $\phi_{ij}(w) = 0$ for all $j
\in {\cal B}_i^k(w)$, where ${\cal B}_i^k(w)$ is the blocked node set of node $i$
relative to session $w$. The meaning and use of the blocked node set were briefly
discussed in the routing message exchange protocol in
Section~\ref{subsec:RoutingVariables}. The exchange protocol also gives every node enough
information to calculate the current $\delta\boldsymbol\phi_i(w)$ prior to each iteration
of \eqref{eq:InterNodeRT}.

For the source node $O(w)$ of a session $w$, the algorithm \eqref{eq:InterNodeRT} is
applied to the vector of routing variables associated with real outgoing links. The
overflow routing variable $\phi_w$ is updated by \be\label{eq:OverflowRT} \phi_w^{k+1} :=
\left[\phi_w^{k} - \kappa_w^k \left(D_w' - \frac{\partial E}{\partial
r_w}\right)\right]^+, \ee for which the projection is onto the feasible set $0 \le \phi_w
\le 1$. The gradient is given by subtracting $\frac{\partial E}{\partial r_w}$, which is
computable after the routing message exchange, from the local measure $D_w'$.

Finally, we come to the algorithm that node $i$ uses to iteratively adjust the intra-node
routing vector $\bs\mu_{ij} \triangleq (\mu_{ij}(q))_{q \in {\cal OC}_i}$ applied to an
outgoing link $(i,j)$. The intra-node routing update is iterated as
\be\label{eq:IntraNodeRT} \boldsymbol\mu_i^{k + 1} := \left[\boldsymbol\mu_i^k -
(T_{ij}^k)^{-1} \cdot
\partial\boldsymbol D_{F_{ij}}^k\right]_{T_{ij}^k}^+.  \ee Here, $\partial\boldsymbol D_{F_{ij}}$ is
the vector of partial derivatives $\partial D_{ij}^q / \partial F_{ij}(q)$ of all $q \in
{\cal Q}_{ij}$, which are purely local measures. The scaling matrix $T_{ij}^k$ is
positive definite. The feasible set that the projection refers to is defined by
\eqref{eq:MuConstraint}.

\subsection{Convergence of Algorithms}

Having described the node-based power control and routing algorithms, we now state the
main convergence result in the following theorem.

\vspace{0.1in}\begin{theorem}\label{thm:AlgorithmConvergence}
Given a feasible spectrum allocation $\{{\cal L}_q\}_{q \in {\cal
Q}}$, let $\{\boldsymbol\eta_i^0\}$, $\{\boldsymbol\rho_i^0\}$,
$\{\bs\phi_i(w)\}$, $\{\bs\mu_i\}$ be any feasible initial
transmission power and routing configuration with finite cost.
Then with appropriate scaling matrices, the update sequences
$\{\boldsymbol\eta_i^k\}_{k=1}^\infty$,
$\{\boldsymbol\rho_i^k\}_{k=1}^\infty$,
$\{\bs\phi_i^k(w)\}_{k=1}^\infty$ and
$\{\bs\mu_i^k\}_{k=1}^\infty$ generated by the
algorithms~\eqref{eq:EtaAlg}, \eqref{eq:RhoAlg},
\eqref{eq:InterNodeRT} and \eqref{eq:IntraNodeRT} converge, i.e.,
$\boldsymbol\eta_i^k \to \boldsymbol\eta_i^*$,
$\boldsymbol\rho_i^k \to \boldsymbol\rho_i^*$, $\bs\phi_i^k(w) \to
\bs\phi_i^*(w)$, and $\bs\mu_i^k \to \bs\mu_i^*$ for all $i \in
{\cal N}$ and $w \in {\cal W}$ as $k \to \infty$. Furthermore, the
limiting configuration $\{\boldsymbol\eta_i^*\}$,
$\{\boldsymbol\rho_i^*\}$, $\{\bs\phi_i^*(w)\}$, $\{\bs\mu_i^*\}$
satisfies the conditions in Theorem~\ref{thm:SuffOptCond}, and is
a jointly optimal solution to the MCRA problem if the link cost
function $D(x,F)$ satisfies the condition in
Lemma~\ref{lma:ConvexCost}.
\end{theorem}\vspace{0.1in}

The proof hinges on the fact that, by using appropriate scaling matrices, every iteration
of the algorithms~\eqref{eq:EtaAlg}, \eqref{eq:RhoAlg}, \eqref{eq:RhoAlg} and
\eqref{eq:InterNodeRT} reduces the total cost of the MCRA problem until the optimality
conditions in Theorem~\ref{thm:SuffOptCond} are achieved.  Finding the appropriate
scaling matrices, however, is a major challenge.  One approach to this problem is to
choose the scaling matrices so that they upper bound the Hessian matrices with respect to
the updated variables. In this way, the algorithms closely approximate Newton's method,
hence enjoying fast rate of convergence while simultaneously guaranteeing convergence
from all initial conditions. This method has been successfully adopted in the power
control and routing algorithms for single-radio wireless networks in
\cite{paper:XY06ISIT1}, and can be generalized to the present context.  Due to the
limited space, however, we skip the details.

It is worth noting that convergence does not depend on any particular order of running
the algorithms at different nodes. At any time, any node can update any set of its local
optimization variables via the corresponding algorithm. All that is required for
convergence is that each node iterates every algorithm until the adjusted variables have
marginal costs satisfying conditions in Theorem~\ref{thm:SuffOptCond}. Finally, we note
that convergence occurs from {\em any initial configuration} with finite cost. These
features are crucial to the applicability of these algorithms to large networks which
lack the ability of scheduling and synchronizing node operations.

Extensive simulations indicate that our algorithms are adaptive to time-varying network
state, including channel fading, network topology, and traffic demand. Because every
iteration of any algorithm always reduces the total cost under the current network
condition, our scheme is able to constantly readjust routing and transmission powers
towards the optimum that slowly shifts over time due to the network change.

\section{Conclusion}

We have developed an integrated cross-layer resource allocation scheme for general
wireless multi-hop networks. To satisfy the fundamental duplexing constraints, our scheme
first finds a feasible spectrum allocation by (1) dividing the whole spectrum into
multiple sub-bands and (2) activating conflict-free links on each sub-band. Compared with
traditional scheduling in time, the spectrum allocation technique has a number of
advantages in operational simplicity and amenability to distributed and asynchronous
implementation. By studying an equivalent combinatorial link-coloring problem, we found
that the minimum number of sub-bands required by a feasible spectrum allocation is given
by a simple function of the chromatic number of the network connectivity graph. The
minimum number grows asymptotically at a logarithmic rate with the chromatic number,
attesting to the good scalability of the spectrum allocation technique and its robustness
to network topology changes. We designed a simple distributed and asynchronous algorithm
by which a feasible spectrum allocation can be constructed given enough sub-bands.

Given a feasible spectrum allocation, we developed an analytical framework and a set of
node-based distributed algorithms for optimally allocating transmission powers and
traffic rates on active links. Such a framework is especially suitable for the design of
wireless networks with frequency selective channels. We provided the conditions that an
optimal power control and routing configuration must satisfy. We then designed a set of
distributed power control and routing algorithms using the scaled gradient projection
method. These algorithms can be iterated at individual nodes with little control
overhead. Finally, we demonstrated that the algorithms asymptotically achieve the optimal
configuration regardless of the initial condition and the order of iterating different
algorithms.

\appendix

\subsection{Proof of Lemma~\ref{lma:CmpltGraph2}}\label{app:ProofLma2}

We provide the proof for the case $g > \lfloor Q/2 \rfloor$. The other case $g < \lfloor
Q/2 \rfloor$ can be seen as a corollary by taking complements of all subsets involved in
the first case.

First notice that since $g > \lfloor Q/2 \rfloor$, there are $K = {Q \choose {g-1}}$
distinct subsets $C_1, \cdots, C_K$ of cardinality $g-1$, where $K \ge k = {Q \choose g}
$, the number of distinct subsets of cardinality $g$. Hence, the claim makes intuitive
sense.

Consider the bipartite graph consisting of $\{B_i\}$ and $\{C_j\}$ where a pair of $B_i$
and $C_j$ are connected if and only if $C_j \subset B_i$. In this case, $C_j$ is said to
be a {\em child} of $B_i$, and $B_i$ is said to be a {\em parent} of $C_j$. It is easy to
see that every $B_i$ has $g$ children and every $C_j$ has $Q - g + 1$ parents. Now the
claim in the lemma is equivalent to the existence of a complete matching (of size $k$) of
$\{B_i\}$ and $\{C_j\}$ in the bipartite graph specified above. Because $g > \lfloor Q/2
\rfloor$ implies that $g \ge \frac{Q+1}{2}$, the degree of any $B_i$, which is $g$, is
greater than or equal to $Q - g + 1$, the degree of any $C_j$. In this case, a complete
matching must exist by a corollary of Hall's theorem (cf. Corollary 13.4 of
\cite{book:Gri99}).

\subsection{Proof of Lemma~\ref{lma:CmpltGraph3}}\label{app:ProofLma3}

Suppose $\{B_i\}_{i \in {\cal N}}$ is an optimal solution and $\left| \bigcup_{i}  B_i
\right| = Q$. If $\{B_i\}$ does not have the property in the lemma, we can always modify
it to $\{B_i^*\}$ which satisfies the property with $Q^* = \left| \bigcup_{i}  B_i^*
\right|$ being less than or equal to $Q$. Hence, the lemma follows. Suppose we have a
feasible solution ${\cal B}$ consisting of a collection of $N$ subsets $\{B_i\}_{i=1}^N$.
We show that based on $\{B_i\}$, we can construct $\widetilde {\cal B} = \{\widetilde
B_i\}_{i=1}^N$ such that $|\widetilde B_i| = \lfloor Q/2 \rfloor$ for all $i =1, \cdots,
N$, where $Q = |\bigcup_{i}  B_i|$.

Define $m = \min_{i} |B_i|$ and $M = \max_{i} |B_i|$. If $m < M$, then $m < \lfloor Q/2
\rfloor$ or $M > \lfloor Q/2 \rfloor$ or both. If $m < \lfloor Q/2 \rfloor$, define
${\cal N}_m \triangleq \{ i \in {\cal N}: |B_i| = m\}$, replace each $B_j$, $j \in {\cal
N}_m$, by subset $B_j'$ with cardinality $m+1$ such that $B_j \subset B_j'$ for all $j
\in {\cal N}_m$, and $B_j' \ne B_k'$ if and only if $B_j \ne B_k$. Such a replacement is
possible by Lemma~\ref{lma:CmpltGraph2}. If $M > \lfloor Q/2 \rfloor$, define ${\cal N}_M
\triangleq \{ i \in {\cal N}: |B_i| = M\}$, replace each $B_j$, $j \in {\cal N}_M$, by
subset $B_j'$ with cardinality $M-1$ such that $B_j' \subset B_j$ for all $j \in {\cal
N}_M$, and $B_j' \ne B_k'$ if and only if $B_j \ne B_k$. Such a replacement can be found
also by Lemma~\ref{lma:CmpltGraph2}. Denote by ${\cal N}'$ the subset of nodes whose
$B_i$ is changed (either expanded or reduced). Because $m < M$, ${\cal N}'$ is always
non-empty. It can be verified that the new collection of subsets $\{\{B_j'\}_{j \in {\cal
N}'}, \{B_i\}_{i \notin {\cal N}'}\}$ is another optimal solution
of~\eqref{eq:GenMinColor1}. If the minimum and maximum cardinalities of the new
collection of subsets are equal, we are done. Otherwise repeat the above procedure until
we obtain an even collection of subsets. The iterations terminate in a finite number of
steps since each iteration strictly reduces the difference between $M$ and $m$.

\subsection{Proof of Theorem~\ref{thm:OptCondition}}\label{app:ProofThm1}

We show that whenever one of conditions~\eqref{eq:OptRouting1}-\eqref{eq:OptPower3} is
violated, the present configuration can be improved upon. We take
condition~\eqref{eq:OptRouting1} as an example. Arguments for the other conditions are
similar. By \eqref{eq:DeltaPhi} and \eqref{eq:RoutingMargCost}, we have

\begin{small}
\[
\delta\phi_{ij}(w) \ge \min_{q \in {\cal Q}_{ij}} \left[ {D_{F_{ij}}^q}' \right] +
\frac{\partial E}{\partial t_j(w)}
\]
\end{small}
and
\begin{small}
\[
\frac{\partial E}{\partial t_i(w)} \ge \min_{j \in {\cal O}_i} \left\{\min_{q \in {\cal
Q}_{ij}} \left[ {D_{F_{ij}}^q}' \right] + \frac{\partial E}{\partial t_j(w)}\right\}.
\]
\end{small}

Thus, condition~\eqref{eq:OptRouting1} holds if and only if for all $j \in {\cal O}_i$
such that $\phi_{ij}(w)
> 0$, $\delta\phi_{ij}(w) \le \min_{q \in {\cal Q}_{ik}} \left[ {D_{F_{ik}}^q}' \right] + \frac{\partial E}{\partial
t_k(w)}$ for all $k \in {\cal O}_i$. Suppose condition~\eqref{eq:OptRouting1} is violated
at $i$ with $t_i(w)>0$, i.e., there exists $j \in {\cal O}_i$ such that $\phi_{ij}(w) >
0$, $\delta\phi_{ij}(w)
> \min_{q \in {\cal Q}_{ik}} \left[ {D_{F_{ik}}^q}' \right] + \frac{\partial E}{\partial
t_k(w)} \triangleq \sigma$ for some $k \in {\cal O}_i$. If $F_{ik} > 0$ and
$\delta\phi_{ik}(w) > \sigma$, there must exist $q \in {\cal Q}_{ik}$ with $\mu_{ik}(q)
> 0$ but ${D_{F_{ik}}^q}' > {D_{F_{ik}}^v}'$ for some other $v \in {\cal Q}_{ik}$. Hence,
the cost of the configuration can be further reduced by shifting a tiny portion from
$\mu_{ik}(q)$ to $\mu_{ik}(v)$ as suggested by \eqref{eq:DelMu}. If $\delta\phi_{ik}(w) =
\sigma$, then $\delta\phi_{ij}(w) > \delta\phi_{ik}(w)$, we can reduce the cost by
shifting a tiny portion from $\phi_{ij}(w)$ to $\phi_{ik}(w)$ in light of
\eqref{eq:DelPhi}. If $F_{ik} = 0$, then we can always make $\delta\phi_{ik}(w) = \sigma$
by setting all $\mu_{ik}(q)$ equal to zero but $\mu_{ik}(v)$ equal to one for one $v$
achieving the minimum in $\sigma$. Notice that reconfiguring $\{\mu_{ik}(q)\}_{q \in
{\cal Q}_{ik}}$ does not change the value of $\sigma$ since by assumption $F_{ik} = 0$.
Then, as before, the total cost can be further reduced by shifting a tiny portion from
$\phi_{ij}(w)$ to $\phi_{ik}(w)$.

\subsection{Proof of Theorem~\ref{thm:SuffOptCond}}\label{app:ProofThm2}

Suppose $\{ \bs \rho_i, \bs\eta_i, \{\bs\phi_i(w)\}_{w\in{\cal W}}, \bs\mu_i \}$
satisfies the conditions in Theorem~\ref{thm:SuffOptCond}. Suppose the configuration
yields log-power variables $\{ \bs S(q) \}_{q \in {\cal Q}}$ and link flow variables
$(F_w)_{w\in{\cal W}}$, $(F_{ij}(q))_{ (i,j)\in{\cal L},~q\in{\cal Q}_{ij}}$. Let $\{
\tilde{\bs \rho}_i, \tilde{\bs\eta}_i, \{\tilde{\bs\phi}_i(w)\}_{w\in{\cal W}},
\tilde{\bs\mu}_i \}$ be another feasible configuration which yields $\{ \tilde{\bs S}(q)
\}$, $(\tilde F_w)_{w\in{\cal W}}$, $(\tilde F_{ij}(q))_{(i,j)\in{\cal E},~q\in{\cal
Q}_{ij}}$. Recall that each $D_{ij}^q$ is jointly convex in $\bs S(q) =
(S_{mn}(q))_{(m,n)\in{\cal L}_q}$ and $F_{ij}(q)$ while each $D_w$ is convex in $F_w$.
Moreover, the feasible sets of log-power variables and flow variables are both convex.
Therefore, the cost difference under the two configurations can be bounded by

\begin{small}
\bea &&\sum_{w \in {\cal W}} \tilde D_w + \sum_{(i,j)\in{\cal L}}\sum_{q\in{\cal Q}_{ij}}
\tilde D_{ij}^q - \sum_{w \in {\cal W}} D_w -
\sum_{(i,j)\in{\cal L}}\sum_{q\in{\cal Q}_{ij}} D_{ij}^q \nonumber \\
&\ge& \sum_{w \in {\cal W}} D_{w}'  (\tilde F_w - F_w) + \sum_{(i,j)\in{\cal
L}}\sum_{q\in{\cal Q}_{ij}} {D_{F_{ij}}^q}'  (\tilde F_{ij}(q) - F_{ij}(q))
\nonumber\\
&& + \sum_{(i,j)\in{\cal L}} \sum_{q \in {\cal Q}_{ij}} \sum_{(m,n)\in{\cal L}_q}
\frac{\partial D_{ij}^q}{\partial S_{mn}(q)}(\tilde S_{mn}(q) -
S_{mn}(q)).\label{eq:ThreeSums} \eea \end{small}

We will show that the RHS is nonnegative. We can re-write and lower bound the first two
summations by the series of equalities and inequalities on the top of the next page.
\begin{figure*}[!t]
\begin{small}
\beas &&\sum_{w \in {\cal W}} D_{F_w}' \cdot (\tilde F_w - F_w) + \sum_{(i,j)\in{\cal
L}}\sum_{q\in{\cal Q}_{ij}}
 {D_{F_{ij}}^q}' \cdot (\tilde F_{ij}(q) - F_{ij}(q)) \\
&\stackrel{(a)}{=}&  \sum_{w \in {\cal W}} D_{F_w}' \cdot \bar r_w (\tilde \phi_w -
\phi_w) + \sum_{w \in {\cal W}} \left\{ \sum_{(i,j)\in{\cal L}}\sum_{q\in{\cal Q}_{ij}}
\left[ {D_{F_{ij}}^q}' \cdot \tilde t_i(w) \tilde\phi_{ij}(w)
\tilde\mu_{ij}(q) \right] - \frac{\partial E}{\partial t_{w}(w)} \cdot t_{w}(w) \right\} \\
&& - \sum_{w \in {\cal W}} \left\{ \sum_{j \ne O(w), D(w)} \frac{\partial E}{\partial
t_{j}(w)} \left[ \tilde
t_j(w) - \sum_{i \ne D(w)} \tilde t_i(w) \tilde\phi_{ij}(w) \right] \right\} \\
&\stackrel{(b)}{=}&  \sum_{w \in {\cal W}} \bar r_w \left\{ \left[ D_{w}' \tilde \phi_w +
(1 - \tilde \phi_w) \sum_{j \in {\cal O}_{O(w)}} \tilde\phi_{ij}(w) \left(
\sum_{q\in{\cal Q}_{ij}} {D_{F_{ij}}^q}' \tilde\mu_{ij}(q) + \frac{\partial E}{\partial
t_{j}(w)} \right) \right]  - \left[ D_{w}'  \phi_w + (1 - \phi_w) \frac{\partial
E}{\partial t_{w}(w)} \right] \right\} \\
&& + \sum_{w \in {\cal W}} \left\{ \sum_{i \ne O(w), D(w)} \tilde t_i(w) \left[ \sum_{j
\in {\cal O}_{i}} \tilde\phi_{ij}(w) \left(  \sum_{q\in{\cal Q}_{ij}} {D_{F_{ij}}^q}'
\tilde\mu_{ij}(q) + \frac{\partial
E}{\partial t_{j}(w)} \right) - \frac{\partial E}{\partial t_{i}(w)} \right] \right\} \\
&\stackrel{(c)}{\ge}&  \sum_{w \in {\cal W}} \bar r_w \left\{ \left[ D_{w}' \tilde \phi_w
+ (1 - \tilde \phi_w) \frac{\partial E}{\partial t_{w}(w)} \right] - \left[ D_{w}' \phi_w
+ (1 - \phi_w) \frac{\partial E}{\partial t_{w}(w)} \right] \right\}  + \sum_{w \in {\cal
W}} \left\{ \sum_{i \ne O(w), D(w)} \tilde t_i(w) \left[ \frac{\partial E}{\partial
t_{i}(w)} - \frac{\partial E}{\partial t_{i}(w)} \right] \right\} \\
&=& \sum_{w \in {\cal W}} \bar r_w  (\tilde\phi_w - \phi_w) \left(D_{F_w}' -
\frac{\partial E}{\partial t_{w}(w)}\right) \ge 0.  \eeas
\end{small}\hrulefill\end{figure*}

For equality (a), we used the relation that
\[
\sum_{w \in {\cal W}}\frac{\partial E}{\partial t_{w}(w)} \cdot t_{w}(w) =
\sum_{(i,j)\in{\cal L}}\sum_{q\in{\cal Q}_{ij}} {D_{F_{ij}}^q}' \cdot F_{ij}(q).
\]
and appended terms $\frac{\partial E}{\partial t_{j}(w)} \left[ \tilde t_j(w) - \sum_{i
\ne D(w)} \tilde t_i(w) \tilde\phi_{ij}(w) \right]$ which are all equal to zero by flow
conservation constraints. Equality (b) is obtained by rewriting $r_w$ and $\tilde r_w$ as
$\bar r_w (1 - \phi_w)$ and $\bar r_w (1 - \tilde\phi_w)$, respectively, and reorganizing
the terms. Condition~\eqref{eq:OptRouting1} implies that
\[
\frac{\partial E}{\partial t_{i}(w)} \le \sum_{j \in {\cal O}_i}\sum_{q\in{\cal Q}_{ij}}
{D_{F_{ij}}^q}' \tilde\mu_{ij}(q) + \frac{\partial E}{\partial t_{j}(w)},
\]
by which we obtain (c). The final inequality follows from
condition~\eqref{eq:OptRouting2}.

Next we lower bound the third summation on the RHS of \eqref{eq:ThreeSums}. To begin
with, we switch the roles of $(m,n)$ and $(i,j)$ in the summation to rewrite it as
\[
\sum_{(i,j)\in{\cal L}} \sum_{q \in {\cal Q}_{ij}} \sum_{(m,n)\in{\cal L}_q}
\frac{\partial D_{mn}^q}{\partial S_{ij}(q)}(\tilde S_{ij}(q) - S_{ij}(q)).
\]
The partial derivative is computed as
\[
\frac{\partial D_{mn}^q}{\partial S_{ij}(q)} = \left\{ \ba{ll}
                                                        {D_{x_{ij}}^q}' x_{ij}(q), \quad
                                                        &\textrm{if~}(i,j)=(m,n),\\
                                                        -{D_{x_{mn}}^q}' x_{mn}(q) \frac{G_{in}^q
                                                        P_{ij}(q)}{IN_{mn}(q)}, &\textrm{otherwise}.
                                                        \ea \right.
\]
Thus, we can expand the summation as shown in the second block on the next page.
\begin{figure*}[!t]
\begin{small}
\bea && \sum_{(i,j)\in{\cal L}} \sum_{q \in {\cal Q}_{ij}} \sum_{(m,n)\in{\cal L}_q}
\frac{\partial D_{mn}^q}{\partial S_{ij}(q)}(\tilde S_{ij}(q) - S_{ij}(q)) \nonumber\\
&=& \sum_{(i,j)\in{\cal L}} \sum_{q \in {\cal Q}_{ij}} \left[
\sum_{\substack{(m,n)\in{\cal L}_q\\(m,n)\ne(i,j)}} -{D_{x_{mn}}^q}' x_{mn}(q)
\frac{G_{in}^q P_{ij}(q)}{IN_{mn}(q)} + {D_{x_{ij}}^q}' x_{ij}(q)\right]
\ln\frac{\tilde P_{ij}(q)}{P_{ij}(q)} \nonumber \\
&=& \sum_{(i,j)\in{\cal L}} \sum_{q \in {\cal Q}_{ij}} \left[ \sum_{(m,n)\in{\cal L}_q}
-{D_{x_{mn}}^q}' x_{mn}(q) \frac{G_{in}^q P_{ij}(q)}{IN_{mn}(q)} + {D_{x_{ij}}^q}'
x_{ij}(q)(1 + x_{ij}(q))\right]
\ln\frac{\tilde P_{ij}(q)}{P_{ij}(q)} \nonumber \\
&=& \sum_{(i,j)\in{\cal L}} \sum_{q \in {\cal Q}_{ij}} \left[ \sum_{n \in {\cal N}}
G_{in}^q MSG_n^q +
\delta\eta_{ij}(q) \right] P_{ij}(q) \ln\frac{\tilde\rho_i(q) \tilde\eta_{ij}(q)}{\rho_i(q) \eta_{ij}(q)} \nonumber \\
&=& \sum_{i \in {\cal N}} \bar P_i \sum_{q \in {\cal OC}_i} \rho_i(q) \sum_{\substack{j
\in {\cal O}_i\\(i,j)\in{\cal L}_q}}\left[\sum_{n \in {\cal N}} G_{in}^q MSG_n^q +
\delta\eta_{ij}(q)\right] \eta_{ij}(q) \ln \frac{\tilde\eta_{ij}(q)}{\eta_{ij}(q)} +
\sum_{i \in {\cal N}} \sum_{q \in {\cal OC}_i} \rho_i(q) \delta\rho_i(q)
\ln\frac{\tilde\rho_i(q)}{\rho_i(q)}. \label{eq:TwoSums} \eea
\end{small}\hrulefill\end{figure*}

For the first summation on the RHS of \eqref{eq:TwoSums}, we use the convention that $0/0
= 1$ and $y / 0 \ge 1$ for all $y \ge 0$. As a consequence, the summand vanishes if
$\tilde\eta_{ij}(q) = \eta_{ij}(q) = 0$. Moreover, for those $(i,j) \in {\cal L}_q$ but
$\eta_{ij}(q) = 0$, we can lower bound the summand by replacing $\delta\eta_{ij}(q)$ with
$\gamma_i(q)$ (cf. condition~\eqref{eq:OptPower3}) to get
\[
\left[\sum_{n \in {\cal N}} G_{in}^q MSG_n^q + \gamma_i(q) \right]\eta_{ij}(q) \ln
\frac{\tilde\eta_{ij}(q)}{\eta_{ij}(q)}.
\]
It is implicit from \eqref{eq:DeltaRho} and \eqref{eq:OptPower3} that $\delta\rho_i(q) =
\bar P_i [\sum_{n \in {\cal N}} G_{in}^q MSG_n^q + \gamma_i(q)]$.  We thus can lower
bound the first summation in~\eqref{eq:TwoSums} by
\beas &&\sum_{i \in {\cal N}} \sum_{q
\in {\cal OC}_i} \rho_i(q) \delta\rho_i(q) \sum_{\substack{j \in {\cal O}_i\\(i,j)\in
{\cal L}_q}}
\eta_{ij}(q) \ln \frac{\tilde\eta_{ij}(q)}{\eta_{ij}(q)} \\
&\ge& \sum_{i \in {\cal N}} \sum_{q \in {\cal OC}_i} \rho_i(q) \delta\rho_i(q)
\sum_{\substack{j \in {\cal O}_i\\(i,j)\in {\cal L}_q}} \eta_{ij}(q) \left(
\frac{\tilde\eta_{ij}(q)}{\eta_{ij}(q)} - 1 \right) = 0. \eeas
The inequality follows
from the relation $\ln x \le x - 1$ and that $\rho_i(q) \delta\rho_i(q) \le 0$ due to
conditions \eqref{eq:OptPower1} and \eqref{eq:OptPower2}.

Now consider the second summation on the RHS of~\eqref{eq:TwoSums}. For each $i \in {\cal
N}$, let the summands with $\rho_i(q) = 0$ be lower bounded by \[ \rho_i(q) \lambda_i
\ln\frac{\tilde\rho_i(q)}{\rho_i(q)}.
\] By condition~\eqref{eq:OptPower1}, summands with $\rho_i(q) > 0$ are equal to \[ \rho_i(q)
\lambda_i \ln\frac{\tilde\rho_i(q)}{\rho_i(q)}.\] Therefore, the whole summation is lower
bounded by \beas &&\sum_{i \in {\cal N}} \sum_{q \in {\cal OC}_i} \rho_i(q) \lambda_i
\ln\frac{\tilde\rho_i(q)}{\rho_i(q)} \\
&\ge& \sum_{i \in {\cal N}} \sum_{q \in {\cal OC}_i} \rho_i(q) \lambda_i \left(
\frac{\tilde\rho_i(q)}{\rho_i(q)} - 1 \right) \\
&=&  \sum_{i \in {\cal N}} \lambda_i \left(\sum_{q \in {\cal OC}_i} \tilde\rho_i(q) -
\sum_{q \in {\cal OC}_i} \rho_i(q) \right) \ge 0. \eeas The second to last inequality is
obtained by noting that $\rho_i(q) \lambda_i \le 0$ due to condition~\eqref{eq:OptPower2}
and the identity $\ln x \le x - 1$. The last inequality follows also from
condition~\eqref{eq:OptPower2}.

Thus, we have shown that the cost incurred by an arbitrary feasible configuration is
greater than or equal to the cost by a configuration that satisfies the conditions in
Theorem~\ref{thm:SuffOptCond}. So the proof is complete.

\bibliography{./main,FreqSchd,WirelessRouting}
\bibliographystyle{ieeetr}

\begin{biography}{Yufang Xi} (S'06) received the B. Sci degree in Electronic Engineering from Tsinghua University, Beijing, China, in
2003, and the M. Sci and M. Phil degrees in Electrical Engineering from Yale University,
New Haven, CT, in 2005 and 2006, respectively. He is currently a Ph.D. candidate at the
Department of Electrical Engineering at Yale.

His research interests include cross-layer optimization and distributed resource
allocation algorithms for wireless networks.
\end{biography}

\begin{biography}{Edmund M. Yeh}
received his B.S. in Electrical Engineering with Distinction from Stanford University in
1994, M.Phil in Engineering from the University of Cambridge in 1995, and Ph.D. in
Electrical Engineering and Computer Science from MIT in 2001.  Since July 2001, he has
been on the faculty at Yale University, New Haven, Connecticut, where he is currently an
Associate Professor of Electrical Engineering and Computer Science.

Dr. Yeh is a recipient of the Army Research Office (ARO) Young Investigator Program (YIP)
Award (2003), the Winston Churchill Scholarship (1994), the National Science Foundation
and Office of Naval Research Fellowships (1994) for graduate study, the Frederick E.
Terman Award from Stanford University (1994) and the Barry M. Goldwater Scholarship from
the United States Congress (1993).  Dr. Yeh is a member of Phi Beta Kappa, Tau Beta Pi,
and IEEE.  He has been visiting faculty at MIT, Princeton University, University of
California at Berkeley, and Swiss Federal Institute of Technology, Lausanne.
\end{biography}

\end{document}